\documentclass[aps,letterpaper,twocolumn,pra,footinbib,floatfix,showpacs]{revtex4}
\usepackage{amsmath,verbatim,graphicx,psfrag,natbib}

\begin{document}
\author{Bhuvanesh Sundar}
\affiliation{Laboratory of Atomic and Solid State Physics, Cornell University, Ithaca NY 14850}
\author{Erich J. Mueller}
\affiliation{Laboratory of Atomic and Solid State Physics, Cornell University, Ithaca NY 14850}
\title{Universal Quantum Computation With Majorana Fermion Edge Modes Through Microwave Spectroscopy Of Quasi-1D Cold Gases In Optical Lattices}
\date{\today}

\begin{abstract}
  We describe how microwave spectroscopy of cold fermions in quasi-1D traps can be used to detect, manipulate, and entangle exotic nonlocal qubits associated with ``Majorana'' edge modes. We present different approaches to generate the $p$-wave superfluidity which is responsible for these topological zero-energy edge modes. We find that the edge modes have clear signatures in the microwave spectrum, and that the line shape distinguishes between the degenerate states of a qubit encoded in these edge modes. Moreover, the microwaves rotate the system in its degenerate ground-state manifold. We use these rotations to implement a set of universal quantum gates, allowing the system to be used as a universal quantum computer.
\end{abstract}

\newcommand{\ha}{\hat{a}}
\newcommand{\hc}{\hat{c}}
\newcommand{\hmaj}{\hg_{\rm{edge}}}
\newcommand{\hg}{\hat{\gamma}}
\newcommand{\hH}{\hat{H}}
\newcommand{\gplus}{|g_+\rangle}
\newcommand{\gminus}{|g_-\rangle}
\newcommand{\gpm}{|g_\pm\rangle}
\newcommand{\+}{^\dagger}

\pacs{67.85.Lm, 03.75.Ss, 03.65.Vf, 03.67.Lx}
\maketitle

\section{Introduction}
In condensed-matter physics, Majorana fermions are exotic collective excitations that possess non-Abelian braiding statistics and are ``topologically protected'' \cite{nonabelionsReview}. They can be used to realize fault-tolerant quantum bits (qubits), an essential ingredient for quantum computing. In this paper we propose a scheme in which microwaves are used to detect Majorana modes in a cold Fermi gas, read out the state of the system, and perform quantum gates.

In a seminal paper Kitaev showed that one-dimensional (1D) $p$-wave superfluids support zero-energy Majorana fermion excitations at their edges \cite{Kitaev2001}. Motivated in part by these arguments, researchers have searched for (and found) evidence of Majorana modes in spin-orbit coupled nanowires with proximity-induced superconductivity \cite{Kouwenhoven2012,Rokhinson2012}. In ultracold atomic systems, researchers are attempting to produce Majorana modes using Raman-induced spin-orbit coupling \cite{Zwierlein2012,Zhang2012}. They have created spin-orbit coupled systems with strongly interacting fermions and measured their radio-frequency spectra \cite{fu2013radio}.
Our proposal builds on these developments. We explain how $p$-wave superfluidity can be induced in quasi-1D cold gases, and study the interactions of these systems with microwaves. In particular, we describe protocols to produce a set of universal quantum gates.

The ground state of a 1D $p$-wave superfluid is doubly degenerate. These two ground states can be used as a qubit. The information encoded in this qubit is nonlocal: These states can be distinguished by simultaneously probing the two boundaries of the system or by measuring particle number parity. Mathematically one describes the system as having a Majorana mode at each edge and attributes the degeneracy to these Majorana degrees of freedom.

Detecting Majorana modes is a major challenge and has attracted wide theoretical and experimental interest \cite{Kouwenhoven2012, Rokhinson2012, Ran2012, Sylvaine2012, MajoranaFermionProbe1, MajoranaFermionProbe2, MajoranaFermionProbe4, MajoranaFermionProbe5, MajoranaFermionProbe6}. In order to use Majorana edge states as qubits, one must also have the capability to perform projective measurements and quantum gate operations in the degenerate ground-state manifold. Here we note that optical, microwave, or radio probes coherently interact with well-separated regions of a cold-gas system and can be used for detecting and manipulating the delocalized quantum information. We show that the system's microwave-absorption spectrum measures the qubit and show that microwaves also rotate the qubit.

We emphasize that while other spectroscopic techniques can detect the existence of Majorana modes, the microwave absorption spectroscopy studied here can do more: It allows us to measure and manipulate the stored quantum information. Further, while other theoretical studies have explored manipulation of this stored information by interactions of Majorana fermions with microwave cavities \cite{MicrowaveCouplingOfMajoranas,schmidt2013majorana}, we do more: We describe methods to produce all the necessary single- and two-qubit quantum gates for universal quantum computation including initialization and measurement and describe a general framework for producing arbitrary rotations of the qubits.

We envision two possible experimental geometries of superfluids that support Majorana fermions. One geometry consists of a 2D array of 1D superfluid wires with weak interwire tunneling. We believe this is the most experimentally accessible geometry for creating Majorana fermions in cold gases. Multiple wires are advantageous because they enhance the measured signal. Moreover, the interwire tunneling stabilizes long-range superfluid order and allows a mean-field treatment of each wire. The other geometry consists of a single wire embedded in a 3D superfluid bath. While the linear response to microwaves in this geometry is the same as that for an array of wires, the dynamics beyond linear response are different in these two cases. The protocols necessary for quantum computations work better in this latter geometry. Further the latter geometry connects more closely to experiments with evidence of Majorana fermions in solid state systems.

This paper is organized as follows. In Sec. \ref{sec:Exptal Setup} we briefly review experimental scenarios to realize $p$-wave superfluidity in arrays of coupled wires and in wires in proximity to a superfluid cloud. In Sec. \ref{sec: Model} we present our theoretical model for these systems. In Sec. \ref{sec:qps} we explore the single-particle excitations in this model. In Sec. \ref{sec:Majorana modes} we specialize to the Majorana modes. In Sec. \ref{sec:Gamma} we calculate the response of the system to a microwave probe. In Secs. \ref{sec:Dynamics}--\ref{sec:gates} we explain how to implement various quantum gates on single qubits. In Sec. \ref{sec:Dynamics} we explore the relative time scales required for performing different gate operations. In Sec. \ref{sec:numerical methods} we discuss our numerical methods for studying the dynamics of the system during gate operations. In Sec. \ref{sec:gates} we discuss quantum gates through microwave illumination of qubits. In Sec. \ref{sec:ancillary bit} we describe an architecture and algorithms to implement all the quantum gates necessary for universal quantum computation in this system. We summarize in Sec. \ref{sec:Summary}.

\section{Experimental setup} \label{sec:Exptal Setup}
We consider two possible geometries of cold neutral fermionic atoms or molecules that will have Majorana fermion excitations at their edges, a 2D array of 1D wires created by a highly anisotropic optical lattice and a 1D potential valley on the surface of an atom chip embedded in a large 3D superfluid cloud. In the first geometry a weak interwire tunneling stabilizes superfluidity in the wires. In the second geometry superfluidity is induced in the wire due to the proximity effect. Below we propose experimental implementations of these two geometries. We discuss methods to create $p$-wave superfluids in these geometries. We supplement these discussions with calculations of experimental parameters in the coupled-wire geometry.

\subsection{Coupled Wires} \label{subsec:coupled wires expt}
First we study a fermionic gas trapped in a 2D array of 1D wires created by a highly anisotropic optical lattice: Hopping in the $y$ and $z$ directions is strongly suppressed compared to that in the $x$ direction. Such an array is readily produced by interfering several lasers \cite{Hulet2010}. $p$-wave superfluidity will be favored if there are strong nearest-neighbor interactions in each wire. Below we review a few methods to generate such interactions. In Appendix \ref{sec:Appendix} we calculate the superfluid gap ($\Delta$) created by these interactions. For our proposals to work, we would like the gap $\Delta$ to be such that $\Delta/k_B$ is on the order of or larger than a few nanokelvins, which is the typical temperature of these gases. For reference, $\hbar\cdot 1\ \rm{MHz} = k_B\cdot 7.64\ \rm{nK}$.

The conceptually simplest way to generate nearest-neighbor (or longer-ranged) interactions involves using neutral molecules such as KRb or LiCs that have large dipole moments \cite{ZollerDipoles}. Interactions between the molecules can be controlled by applying an electric field, which adjusts the alignment of their dipole moments. The dipole moments of these molecules are typically of the order of 10 D, which creates interactions of strengths in the kHz range between nearest neighbors on a lattice with lattice spacing of a few $\mu$m. As shown in Appendix \ref{sec:Appendix}, the superfluid gap will then be in the kHz range.

A second method to create nearest-neighbor interactions uses spin-orbit coupled atoms in the lower helicity state. Here one shines two counter propagating lasers on two spin species of fermionic atoms such as $^{40}$K or $^6$Li. The resulting dressed atomic levels are described by a Hamiltonian analogous to spin-orbit coupled electrons in a semiconducting wire. When projected into the lower dressed band, effective separable long-range atomic interactions emerge \cite{Zhang2012}. Current experiments using $^{40}$K \cite{Zhang2012} achieve spin-orbit coupling strengths and Zeeman splitting between the dressed states in the kHz range. In Appendix \ref{sec:Appendix} we show that the resulting superfluid gap will be of order kHz.

A third method to create nearest-neighbor interactions is to engineer significant overlap of Wannier functions at adjacent lattice sites. One such example is to use spin-dependent lattices, outlined in \cite{Sylvaine2012}. We show in Appendix \ref{sec:Appendix} that the resulting superfluid gap would be of order MHz.

All of the methods outlined above create effectively spinless fermions in one band ``a'' which have interactions that extend beyond a single site. While these techniques generate interactions of various ranges, the essential physics is captured by a model with only nearest-neighbor interactions. We work with this simplified model.

\subsection{Proximity-Induced Superfluid} \label{subsec:proximity-induced superfluid expt}
We also consider a geometry where a single wire is immersed in a superfluid cloud. Superfluidity is induced in the wire due to the proximity effect. This is the geometry realized in solid-state experiments, where semiconductor wires are in contact with a superconductor \cite{Kouwenhoven2012}. One way to realize this geometry in a gas of cold atoms is to put an $s$-wave superfluid of $^6$Li or $^{40}$K near an atom chip \cite{AtomChip2001}. Current-carrying wires on the atom chip create a 1D potential valley for the atoms. We envision a bimodal population, with a large 3D cloud surrounding a tightly confined 1D gas. We artificially induce spin-orbit coupling in the 1D gas by shining Raman lasers along the potential valley. By keeping the beam waist small, the Raman lasers have minimal effect on the 3D cloud. If desired, an optical lattice can be added to the potential valley. The superfluid cloud acts as a bath of Cooper pairs which tunnel into the 1D wire; hence, a superfluid order parameter is induced in the chain of atoms. The artificially induced spin-orbit coupling projects these pairs onto dressed bands. In the lower dressed band the induced order parameter has $p$-wave character.

\subsection{Trap and Probe}
Most experiments in both geometries are performed in the presence of harmonic potentials. As explicitly shown by Wei and Mueller \cite{Ran2012} for the case of spin-orbit coupled atoms, the physics of Majorana fermions in a harmonic trap is identical to that in an infinite square well. Experimentally infinite square wells can be engineered with ``tube beams'' and ``sheet beams'' \cite{FlatCondensate2013}. For simplicity, we predominantly model the confinement as an infinite square well. In Sec. \ref{subsec:harmonic} we also include calculations in a harmonic trap.

To probe the Majorana modes in the 1D superfluid, we propose driving a transition to another atomic state ``c'' by shining electromagnetic waves on the system. We consider the case of microwaves or radiowaves, where the wavelength is larger than the system size ($0.1-1$mm). If the bandwidth of the Bloch waves in the ``c'' states is small compared to $\Delta$, we find that the spectral signature of the Majorana modes is well separated from the bulk modes, and the spectral line shape can distinguish which of the two degenerate ground states is present. Researchers have been able to create spin-dependent lattices with different bandwidths for different hyperfine states of bosonic atoms \cite{demarcoThermometry}. We anticipate that similar techniques can be used to independently control the bandwidths of the fermionic ``a'' and ``c'' states in our system as well.

\section{Theoretical model} \label{sec: Model}
The two geometries discussed in the previous section are modeled by slightly different effective Hamiltonians. The cause of the difference is the mechanism inducing superfluidity in the wires. In the coupled-wire geometry, the superfluid order parameter $\Delta$ is determined self-consistently from the properties of the wires, while in the case of a proximity-induced superfluid, $\Delta$ is imposed by the surrounding bath. Below we present the Hamiltonian governing our system in these two cases.

\subsection{Coupled Wires} \label{subsec:coupled wires model}
The ensemble of wires described in Sec. \ref{subsec:coupled wires expt} is modeled by the effective Hamiltonian
\begin{equation}
 \hH = \hH_a + \hH_c + \hH_{\rm{MW}} + \hH_{\rm{iw}}.
\end{equation}
The atoms in the ``a'' states are described by a tight-binding model,
\begin{equation}\begin{split} \label{eqn:full Hsys}
 \hH_a = \sum_i \sum_{j=1}^{N-1} -J\left(\ha_j^{(i)\dagger}\ha_{j+1}^{(i)} +\ha_{j+1}^{(i)\dagger}\ha_j^{(i)}\right) \\
  + V\ha_j^{(i)\dagger}\ha_{j+1}^{(i)\dagger} \ha_{j+1}^{(i)}\ha_j^{(i)} - \sum_i\sum_{j=1}^N \mu\ha_j^{(i)\dagger}\ha_j^{(i)},
\end{split}\end{equation}
where $J$ is the hopping amplitude, $V$ is the nearest-neighbor interaction strength, $\mu$ is the chemical potential, and $\ha_j^{(i)\dagger}$ and $\ha_j^{(i)}$ create and annihilate spinless fermions in the ``a'' state at site $j$ in the $i^{\rm{th}}$ wire. We model the interwire coupling via
\begin{equation}\begin{split}
\hH_{\rm{iw}} = \sum_{<ii'>}\sum_{j=1}^N-J'\left(\ha_j^{(i)\dagger} \ha_j^{(i')} + \ha_j^{(i')\dagger}\ha_j^{(i)}\right)\\
 + V'\ha_j^{(i)\dagger}\ha_j^{(i')\dagger}\ha_j^{(i')}\ha_j^{(i)} \label{eqn:Hiw}
\end{split}\end{equation}
where $J'$ and $V'$ parametrize the interwire hopping and interactions, respectively. All adjacent pairs of wires $\langle ii'\rangle$ are summed over. In the limit of $J'<<J$ and $V'<<V$, $\hH_{\rm{iw}}$ has only a perturbative role in this mean-field theory and can be neglected. Under a mean-field approximation, each wire is then equivalent, and we can drop the index labeling the wires from our operators. We emphasize, however, that the presence of this interwire hopping is important as it stabilizes long-range order and justifies our mean-field approximation. The mean-field Hamiltonian describing the ``a'' atoms or molecules is then
\begin{equation}\begin{split} \label{eqn:HMF}
 \hH_{\rm{MF}} = E_0 + \sum_{j=1}^{N-1}\Big[-J\left(\ha_j\+\ha_{j+1} +\ha_{j+1}\+\ha_j\right)\\
   -\left(\Delta_j\ha_j\+\ha_{j+1}\+ + \Delta_j^*\ha_{j+1}\ha_j\right)\Big] - \sum_{j=1}^N \mu\ha_j\+\ha_j,
\end{split}\end{equation}
where the local order parameter $\Delta_j$ is defined self-consistently as
\begin{equation}
\Delta_j = -V\langle\ha_{j+1}\ha_j\rangle,
\label{eqn:self-consistency}
\end{equation}
and $E_0$ is an irrelevant energy shift. $\Delta_j$ can always be chosen to be real and positive by a local gauge transformation of the creation and annihilation operators of the ``a'' atoms or molecules. Moreover, one can also take $\Delta_j$ to be symmetric under reflection, $\Delta_j = \Delta_{N-j}$.

Microwaves can change the internal state of the atoms or molecules, introducing a term in the Hamiltonian of the form
\begin{equation}
 \hH_{\rm{MW}} = \Omega \sum_{j=1}^N \left(\hat{c}_j^\dagger\hat{a}_j e^{-i\omega t} + \hat{a}_j^\dagger\hat{c}_j e^{i\omega t}\right).
 \label{eqn:HMW}
\end{equation}
The operators $\hat{c}_j$ and $\hat{c}_j\+$ correspond to excited atoms or molecules. These excited particles generally feel a different lattice, and are described by a tight-binding model
\begin{equation}
 \hH_c = \sum_j -J_c\left(\hat{c}_j^\dagger\hat{c}_{j+1} +\hat{c}_{j+1}^\dagger\hat{c}_j\right) -\mu_c \hat{c}_j^\dagger\hat{c}_j,
 \label{eqn:Hprobe}
\end{equation}
where $J_c$ is the hopping amplitude and $\mu_c$ is the chemical potential. The frequency $\omega$ in Eq. (\ref{eqn:HMW}) is related to the real frequency of electromagnetic waves via
\begin{equation}
 \omega = \omega_{\rm{physical}} - \delta\epsilon + \mu - \mu_c,
 \label{eqn:frequency relation}
\end{equation}
where $\omega_{\rm{physical}}$ is the real electromagnetic frequency and $\delta\epsilon$ is the absorption frequency in free space.

\subsection{Proximity-Induced Superfluid} \label{subsec:proximity-induced superfluid model}
A proximity-induced superfluid will be governed by the effective Hamiltonian
\begin{equation}
 \hH = \hH_a + \hH_c + \hH_{\rm{MW}}. \label{eqn:Hfull}
\end{equation}
The ``a'' states are described by a tight-binding model
\begin{equation}\begin{split}
 \hH_a = \sum_{j=1}^{N-1}\Big[-J\left(\ha_j\+\ha_{j+1} +\ha_{j+1}\+\ha_j\right) \\ - \left(\Delta_j\ha_j\+\ha_{j+1}\+ + \Delta_j^*\ha_{j+1}\ha_j\right)\Big] - \sum_{j=1}^N \mu\ha_j\+\ha_j, \label{eqn:HMF proximity}
\end{split}\end{equation}
where $J$ is the hopping amplitude, $\Delta_j$ is the order parameter induced at site $j$ from the bath, $\mu$ is the chemical potential, and $\ha_j\+$ and $\ha_j$ create and annihilate spinless fermions in the ``a'' state at site $j$ in the wire. $\hH_{\rm{MW}}$ and $\hH_c$ are given by Eqs. (\ref{eqn:HMW}) and (\ref{eqn:Hprobe}). The model for proximity-induced superfluids is simpler because as long as the bath is large enough, $\Delta_j$ are constant parameters and need not be self-consistently determined from the properties of the wire.

\subsection{Other Considerations}
In this study we analyze the case where interactions between ``a'' and ``c'' atoms or molecules can be neglected. For example, this would be the case for spin-orbit coupled $^{40}$K near a Feshbach resonance. Such final-state interactions can be modeled by using the techniques from \cite{Basu2008}.

\section{Bogoliubov excitations} \label{sec:qps}
In this section, we analyze the quasiparticles of the mean-field theory. We give particular emphasis to their properties under reflection and derive relations that will be useful in the next section. We explore the local density of states associated with our mean-field model in Eqs. (\ref{eqn:HMF}) and (\ref{eqn:HMF proximity}) and verify that a zero-energy mode appears at the boundaries. We include a brief qualitative discussion of the full coupled-wire model with the interwire coupling at the end of this section.

Equations (\ref{eqn:HMF}) and (\ref{eqn:HMF proximity}) can be rewritten as
\begin{equation}
 \hH_{\rm{MF}} = E_0' + \sum_\nu E_\nu \hg_\nu^\dagger\hg_\nu,
 \label{eqn:Bogs}
\end{equation}
where $E_0'$ is an irrelevant energy shift. The creation operator for the quasiparticles, which are superpositions of particles and holes, are of the form
\begin{equation}
 \hat\gamma_\nu\+ = \sum_{j=1}^N u_\nu(j)\ha_j\+ + v_\nu(j)\ha_j,
 \label{eqn:u,v}
\end{equation}
where $N$ is the number of sites. Since all the parameters in our mean-field model are real, we can always choose $u_\nu(j)$ and $v_\nu(j)$ to be real.

The coherence factors $u_\nu(j)$ and $v_\nu(j)$ at different sites are not completely independent of each other. We consider an operator that performs a simultaneous reflection and a global gauge transformation,
\begin{equation}
 \hat{Q} \ha_j \hat{Q}\+ = i\ha_{N+1-j}. \label{eqn:symm}
\end{equation}
Under the assumption that $\Delta_j=\Delta_{N-j}$, this operator represents a symmetry of the mean-field Hamiltonian, $[\hat{Q},\hH_{\rm{MF}}]=0$. Barring any degeneracies, this symmetry implies that $[\hat{Q}, \hat\gamma_\nu\+\hat\gamma_\nu]=0\ \forall\nu$. Consequently, $\hat{Q}$ acts a gauge transformation on the quasiparticle operators, $\hat{Q} \hat\gamma_\nu\+ \hat{Q}\+ = \lambda_\nu\hat\gamma_\nu\+$. Direct computation starting from Eq. (\ref{eqn:symm}) yields $\hat{Q}^2\hat\gamma_\nu\+\hat{Q}^{\dagger 2} = -\hat\gamma_\nu\+$. Hence, $\lambda_\nu = \pm i$. When $\lambda_\nu = -i$, $u_\nu(j)$ is symmetric under reflection and $v_\nu(j)$ is antisymmetric under reflection: $u_\nu(N+1-j) = u_\nu(j),\ v_\nu(N+1-j) = -v_\nu(j)$. In this case the quasiparticle creation operators have the form
\begin{equation}
 \hat\gamma_\nu^{(s)\dagger} = \sum_j f_\nu^{(s)}(j)\frac{\ha_j\++\ha_j+\ha_{N+1-j}\+-\ha_{N+1-j}}{2},
\end{equation}
where $f_\nu^{(s)}(j) = u_\nu(j)+v_\nu(j)$. When $\lambda_\nu = i$, $u_\nu(j)$ is antisymmetric under reflection and $v_\nu(j)$ is symmetric under reflection:  $u_\nu(N+1-j) = -u_\nu(j),\ v_\nu(N+1-j) = v_\nu(j)$. In this case the quasiparticle creation operators have the form
\begin{equation}
 \hat\gamma_\nu^{(a)\dagger} = \sum_j f_\nu^{(a)}(j)\frac{\ha_j+\ha_j\++\ha_{N+1-j}-\ha_{N+1-j}\+}{2},
\end{equation}
where again $f_\nu^{(a)}(j) = u_\nu(j)+v_\nu(j)$. These modified forms will be helpful in our analysis of Majorana modes in Sec. \ref{sec:Majorana modes}.

\begin{figure}[htbp]
   \includegraphics[width=0.9\columnwidth]{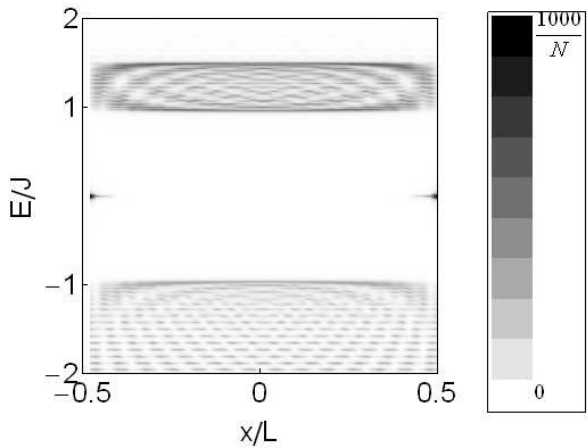}
   \caption{Local single-particle density of states for the mean-field model in Eq. (\ref{eqn:HMF proximity}) with parameters $\Delta_j = \mu = 0.25J, N = 50$. The local density of states is a sum of Dirac $\delta$ functions. Darker shades represent higher amplitudes of the Dirac $\delta$ functions. We can observe bulk quasiparticles with $E_\nu>0$ and zero-energy ($E_\nu=0$) excitations at the edges. These zero-energy modes are Majorana modes.}
   \label{fig:LDOS}
\end{figure}

The local density of states  in the superfluid at energy $E$ and position $j$ is given by
\begin{equation}\begin{split} \label{eqn:spectral fn}
 A(j,E) = 2Im\left(\langle\ha_j\frac{1}{E-\hH}\ha_j\+\rangle + \langle\ha_j\+\frac{1}{E+\hH}\ha_j\rangle\right)\\
 = 2\pi\sum_\nu \left[|u_\nu(j)|^2\delta(E-E_\nu) + |v_\nu(j)|^2\delta(E+E_\nu)\right],
\end{split}\end{equation}
where $\delta$ is the Dirac $\delta$ function, and all quasiparticle indices $\nu$ have been summed over. We illustrate this density of states in Fig. \ref{fig:LDOS}, marking a point in the position-energy plane at the locations of the Dirac $\delta$ functions, with darker points representing higher amplitudes of the $\delta$ functions. The pair of zero-energy peaks visible in Fig. \ref{fig:LDOS} are Majorana modes. These Majorana modes are zero-energy solutions to the Bogoliubov-deGennes equations, that are exponentially localized at the two ends \cite{Kitaev2001}. They always occur in pairs at opposite ends of the 1D superfluid and are always found if $|\mu|<2J$, where the superfluid is in a topologically non-trivial phase.

To explore the robustness of these spectra, we also carried out the analogous calculation using a multi-wire mean-field theory,
\begin{equation}\begin{array}{rc}
 \hat{H}_{\rm{MWMF}} = &\sum_i\sum_{j=1}^N [-J \left( \ha_j^{(i)\dagger}\ha_{j+1}^{(i)} + \ha_{j+1}^{(i)\dagger}\ha_j^{(i)}\right) \\
 & -\left( \Delta_j\ha_j^{(i)\dagger}\ha_{j+1}^{(i)\dagger} + \Delta_j^*\ha_{j+1}^{(i)}\ha_j^{(i)} \right) - \mu \ha_j^{(i)\dagger}\ha_j^{(i)}] \\
 & -J' \sum_{\langle ii'\rangle}\sum_{j=1}^N \left( \ha_j^{(i)\dagger}\ha_j^{(i+1)} + \ha_j^{(i+1)\dagger}\ha_j^{(i)}\right)
\end{array}\end{equation}
with $J$ and $\Delta_j$ comparable in magnitude and $\mu=0$. The interwire coupling $J'$ has very little effect on the spectrum, even when it is a sizable fraction of $J$. In particular, the edge modes stay at zero energy. Hence, our assumption that $\hat{H}_{\rm{iw}}$ only plays a perturbative role is valid, and it is sufficient to work with our mean-field model in Eqs. (\ref{eqn:HMF}) and (\ref{eqn:HMF proximity}) hereafter.

\section{Majorana modes} \label{sec:Majorana modes}
The $E_\nu=0$ quasiparticle in Eq. (\ref{eqn:Bogs}) is composed of two Majorana modes. Since this excitation is localized at the edges, we denote its creation operator by $\hmaj\+$. There is an ambiguity in the definition of $\hmaj\+$, as the canonical transformation $\hmaj\rightarrow\hmaj\+, \hmaj\+\rightarrow\hmaj$ is a symmetry of the system. Somewhat arbitrarily, we define $\hmaj\+$ by its properties under $\hat Q$, $\hat Q\hmaj\+\hat{Q}\+ = -i\hmaj\+$. With this choice,
\begin{equation}
 \hmaj\+ = \sum_j f_0(j)\left(\frac{\ha_j\++\ha_j}{2}+i\frac{\ha_{N+1-j}\+-\ha_{N+1-j}}{2i}\right). \label{eqn:Majorana general}
\end{equation}

The coherence factors $f_0(j)$ are real and satisfy
\begin{equation}\begin{split}
 (J-\Delta_{j-1})f_0(j-1) + (J+\Delta_j)f_0(j+1) + \mu f_0(j) = 0,\\ 1<j<N. \label{eqn:f}
\end{split}\end{equation}
For the proximity-induced case with uniform $\Delta_j=\Delta$, the above equation can be solved explicitly \cite{Kitaev2001}. The resulting solution has the form
\begin{equation}\begin{array}{rllcl}
 f_0(j) &=& \alpha(x_+^j-x_-^j),\ &\rm{if}&\ x_+,x_-\neq0\ \rm{and}\ \frac{1}{x_+},\frac{1}{x_-}\neq0, \\
 f_0(j) &=& \delta_{j1},\ &\rm{if}&\ x_+=x_-=0, \\
 f_0(j) &=& \delta_{jN},\ &\rm{if}&\ \frac{1}{x_+}=\frac{1}{x_-}=0, \label{eqn:f0 solution}
\end{array}\end{equation}
where
\begin{equation}
 x_\pm = \frac{\mu}{2(J+\Delta)} \pm \sqrt{\left(\frac{\mu}{2(J+\Delta)}\right)^2+\frac{\Delta-J}{\Delta+J}},
\end{equation}
$\alpha$ is a normalization constant, and $\delta$ is the Kronecker $\delta$. A derivation of Eq. (\ref{eqn:f0 solution}) is provided in Appendix \ref{sec:Appendix2}. The coherence factors $f_0(j)$ exponentially decay away from the boundaries. They are sharply peaked at $j=1$ if $\Delta>0$, and peaked at $j=N$ if $\Delta<0$. Numerical solutions for the coupled-wire case where $\Delta_j$ are determined self-consistently usually result in nearly uniform $\Delta_j=\Delta$. Therefore the expressions in Eq. (\ref{eqn:f0 solution}) are qualitatively applicable to the coupled-wire case as well.

As explained by Kitaev \cite{Kitaev2001}, the physics is particularly simple if
\begin{equation}
 \mu=0, V=-4J, \label{eqn:trivial case}
\end{equation}
parameters which yield uniform $\Delta_j=J$. For these parameters, the energy gap for bulk Bogoliubov excitations is $2J$ and their bandwidth is zero. The coherence factors are $f_0(j)=\delta_{j1}$. While the bulk of our calculations are performed for generic parameters, this simple limit in Eq. (\ref{eqn:trivial case}) is useful for qualitatively understanding the results.

Since $\hmaj\+$ creates an excitation of zero energy, the mean-field models in Eqs.(\ref{eqn:HMF}) and (\ref{eqn:HMF proximity}) have two degenerate ground states, $\gminus$ and $\gplus$. We identify $\gminus$ with the quasiparticle vacuum, characterized by
\begin{equation}
 \hg_\nu\gminus = 0\ \forall\nu,
\end{equation}
and then define
\begin{equation}
 \gplus = \hmaj\+\gminus.
\end{equation}
$\gplus$ and $\gminus$ are eigenstates of the particle number parity operator $\hat{P}=(-1)^{\sum_j\ha_j\+\ha_j}$. For the choice of $\hmaj\+$ we made in Eq. (\ref{eqn:Majorana general}), $\hat{P}\gplus=-\gplus$ and $\hat{P}\gminus=\gminus$. We use the states $\gplus$ and $\gminus$ to encode physical qubits.

In the case of multiple wires, we assume that all the wires are in the same state. The statement that ``the'' qubit is in the state $\gplus$ denotes that all the wires are in $\gplus$.

In the next section we show how to projectively measure physical qubits in the basis of $\gplus$ and $\gminus$. We find in Sec. \ref{sec:gates} that to perform universal quantum computation, we need a more sophisticated architecture in which logical qubits are constructed from more than one physical qubit. Manipulations of the logical qubits consist of manipulations of the component physical qubits. In Secs. \ref{sec:Dynamics}--\ref{sec:gates}, we describe algorithms for implementing quantum gates on physical qubits and the logic behind them. Finally, in Sec. \ref{sec:ancillary bit} we show how to perform universal quantum computation using logical qubits constructed from physical qubits.

\section{Absorption spectrum} \label{sec:Gamma}
The superfluid's electromagnetic absorption spectrum is given by
\begin{equation}
 \Gamma(|i\rangle,\omega) = \sum_f \frac{2\pi}{\hbar}|\langle f|\hH_{\rm{MW}}|i\rangle|^2\delta(\omega-(E_f-E_i)),
 \label{eqn:FGR}
\end{equation}
where $|i\rangle$ is the initial state and $|f\rangle$ are the final states, and $E_i$ and $E_f$ are their respective energies. The final states $|f\rangle$ have one quasiparticle of energy $E_\nu$ and one atom in the excited state, created by
\begin{equation}
 \hc_k\+ = \sum_j \psi_k(j) \hc_j\+, \label{eqn:psik}
\end{equation}
where $\hc_k\+$ diagonalize the tight-binding Hamiltonian in Eq. (\ref{eqn:Hprobe}) for the excited atoms:
\begin{equation}
 \hH_c = \sum_k\epsilon_k\hc_k\+\hc_k.
\end{equation}
In the case of a translationally invariant system, $k$ would label momentum. We consider more generic cases here, and let $k$ simply be a label of the excited states. We denote the energies of the lowest and highest energy ``c'' states as min($\epsilon_k$) and max($\epsilon_k$) and the bottom and top of the bulk ``a'' Bogoliubov spectrum as min($E_\nu$) and max($E_\nu$), where we take $E_\nu>0$; i.e, we exclude the zero-energy edge state in defining these ranges. The bulk excitations contribute to the absorption spectrum in Eq. (\ref{eqn:FGR}) for min$(\epsilon_k)+$min$(E_\nu)<\omega<$max$(\epsilon_k)$+max$(E_{\nu})$. The edge modes contribute to the spectral weight in the range min$(\epsilon_k)<\omega<$max$(\epsilon_k)$. We consider the case where these spectral features are well separated and restrict ourselves to the edge spectrum.

In terms of the wave functions $\psi_k(j)$, Eq. (\ref{eqn:FGR}) simplifies to
\begin{equation}\begin{split}
 \Gamma(\gpm,\omega) = \frac{2\pi|\Omega|^2}{\hbar}\sum_k\delta(\omega-\epsilon_k)\times\\ \left|\sum_jf_0(j)\left(\psi_k(j)\pm\psi_k(N+1-j)\right)\right|^2,
 \label{eqn:Gamma+-}
\end{split}\end{equation}
where $f_0(j)$ are the coherence factors for the edge state in Eq. (\ref{eqn:Majorana general}), $\delta$ is the Dirac $\delta$ function, and all excitation indices $k$ and sites $j$ have been summed over. In the simple case where $\mu=0,\Delta_j=J$, Eq. (\ref{eqn:Gamma+-}) further simplifies to
\begin{equation}
 \Gamma(\gpm,\omega) = \frac{2\pi|\Omega|^2}{\hbar}\sum_k |\psi_k(1)\pm\psi_k(N)|^2\delta(\omega-\epsilon_k).
\end{equation}
Equation (\ref{eqn:Gamma+-}) tell us that the amplitudes of electromagnetic absorption at the two ends of the wire interfere with each other, and the phase associated with absorption at the ends of the wire is different in the states $\gplus$ and $\gminus$. This generically leads to different spectral weights $\Gamma(\gplus,\omega)$ and $\Gamma(\gminus,\omega)$, and enables us to measure the state of the qubit. For an arbitrary superposition of the two states, the spectral weight is given by
\begin{equation}
 \Gamma(\alpha\gplus + \beta\gminus,\omega) = |\alpha|^2\Gamma(\gplus,\omega) + |\beta|^2\Gamma(\gminus,\omega).
\end{equation}

The wave functions $\psi_k(j)$ depend on the how the ``c'' atoms are confined. In the following, we discuss a range of boundary conditions on the ``c'' atoms.

\subsection{Periodic Boundary Conditions} \label{subsec:periodic}
While difficult to implement experimentally, the simplest boundary conditions for theoretical study are periodic. Specifically, we consider a case where the ``a'' atoms are trapped by an optical lattice in a ring geometry consisting of $N$ sites with lattice spacing $a$. A potential barrier (for example, generated by a blue-detuned laser) prevents hopping along one bond, providing edges to the system. We imagine that the ``c'' atoms are also confined to a ring, but do not see the barrier.

In this geometry, the analysis is simple as the eigenstates in Eq. (\ref{eqn:psik}) are plane waves with wave functions
\begin{equation}
 \psi_k(j) = \frac{e^{i kr_j}}{\sqrt{N}}, \label{eqn:psik periodic}
\end{equation}
where $r_j$ is the position of the site labeled by $j$, and the quantized momenta $k$ obey $e^{i Nka}=1$. The spectral weights are then given by
\begin{equation}\begin{split}
 \Gamma(\gpm,\omega) = \frac{2\pi|\Omega|^2}{N\hbar}\sum_k\delta(\omega-\epsilon_k)\\ \times\left|\sum_jf_0(j)\left(e^{i kr_j} \pm e^{i ka}e^{-i kr_j}\right)\right|^2,
\end{split}\end{equation}
where $\epsilon_k = -2J_c\cos ka-\mu_c$, $J_c$ is the hopping amplitude of the excited atoms, and $\mu_c$ is the chemical potential. In this case $\gplus$ and $\gminus$ have clearly distinguishable spectra. For example, at microwave frequency $\omega=-2J_c-\mu_c$, only the uniform wave function ($k=0$) contributes to the spectral weight, and $\Gamma_-(-2J_c-\mu_c)=0$. Similarly, $\Gamma_+(2J_c-\mu_c)=0$.

\begin{figure}[htbp]
   \unitlength=1in
   \psfragscanon
   \psfrag{x}[][]{\begin{picture}(0,0)
    \put(-0.5,-0.1){\makebox(0,0)[l]{
    $\frac{\omega+\mu_c}{J_c}$
   }}\end{picture}}
   \psfrag{y}[][]{\begin{picture}(0,0)
    \put(0.2,0){\makebox(0,0){
    $\frac{2J_c\hbar}{|\Omega|^2}\Gamma$
   }}\end{picture}}
   \psfrag{a}[][]{\begin{picture}(0,0) \put(0,-1.3){\makebox(0,0){(a)}} \end{picture}}
   \psfrag{b}[][]{\begin{picture}(0,0) \put(0,-1.3){\makebox(0,0){(b)}} \end{picture}}
   \psfragscanoff
   \includegraphics[width=0.5\columnwidth]{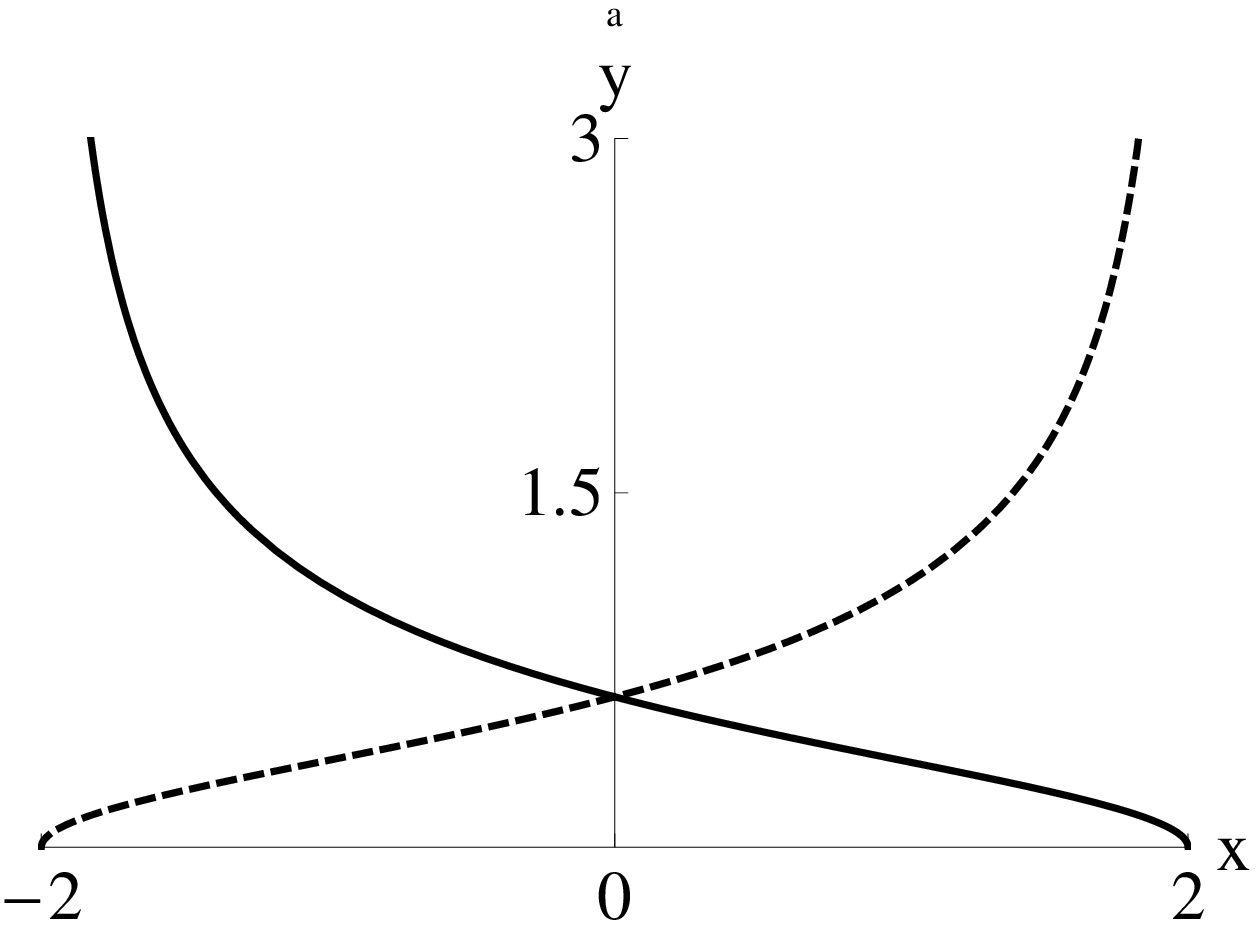}\includegraphics[width=0.5\columnwidth]{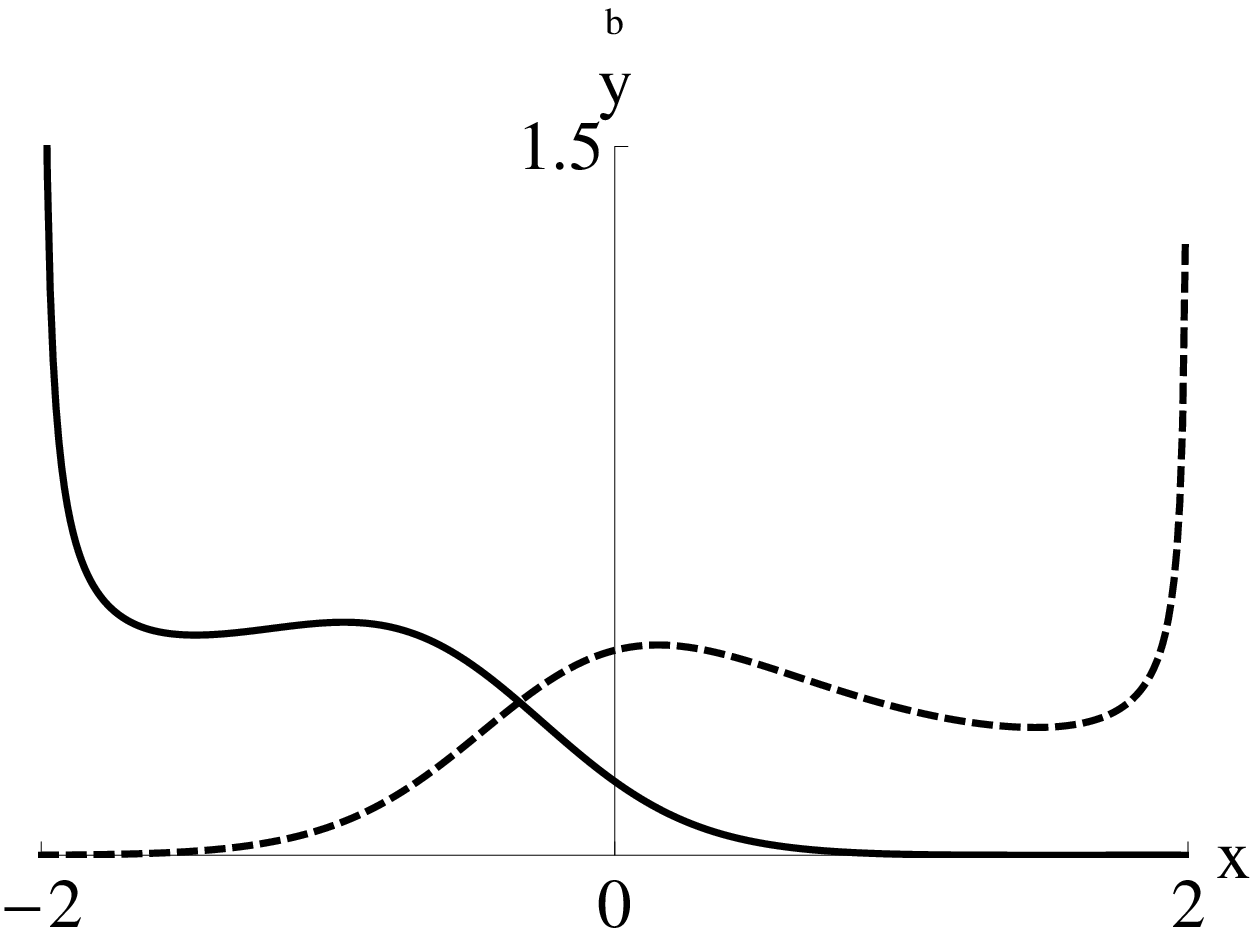}
   \caption{The electromagnetic absorption spectrum when atoms are excited from the ``a'' state to the ``c'' state with periodic boundary conditions. The solid curve is the absorption spectrum of $\gplus$ and the dashed curve of $\gminus$. Parameters chosen in (a) are $\Delta_j=J,\mu=0, N=30$, and in (b) are $\Delta_j=J/2,\mu=J/4, N=30$. The parameters used in (b) are not a self-consistent solution of the mean-field approximation [Eq. (\ref{eqn:f})]. If instead we find $\Delta_j$ self-consistently, the spectra do not change significantly.}
  \label{fig:GammaPeriodic}
\end{figure}

In the special case of Eq. (\ref{eqn:trivial case}), where we took $\mu=0$ and $\Delta_j=J$,
\begin{equation}
 \Gamma(\gpm,\omega) = \frac{-2|\Omega|^2}{J_c\hbar} \frac{-2J_c\pm(\omega+\mu_c)}{\sqrt{(2J_c)^2-(\omega+\mu_c)^2}}.
\end{equation}
This spectrum is plotted in Fig. \ref{fig:GammaPeriodic}(a). For this special case, one sees that $\Gamma_+(\omega)$ is monotonically decreasing while $\Gamma_-(\omega)$ is monotonically increasing. This behavior is special to these parameters, and, as shown in Fig. \ref{fig:GammaPeriodic}(b), the spectra could, in general, have a richer structure.

\subsection{Infinite Square Well} \label{subsec:open trapped}
\begin{figure}[htbp]
   \unitlength=1in
   \psfragscanon
   \psfrag{xaxis}[][]{\begin{picture}(0,0)
    \put(-0.47,-0.12){\makebox(0,0)[l]{
    $\frac{\omega+\mu_c}{J_c}$
   }}\end{picture}}
   \psfrag{y}[][]{\begin{picture}(0,0)
    \put(0.2,0){\makebox(0,0){
    $\frac{2J_c\hbar}{|\Omega|^2}\Gamma$
   }}\end{picture}}
   \psfrag{a}[][]{\begin{picture}(0,0) \put(0,-1.23){\makebox(0,0){(a)}} \end{picture}}
   \psfrag{b}[][]{\begin{picture}(0,0) \put(0,-1.23){\makebox(0,0){(b)}} \end{picture}}
   \psfragscanoff
 \includegraphics[width=0.5\columnwidth]{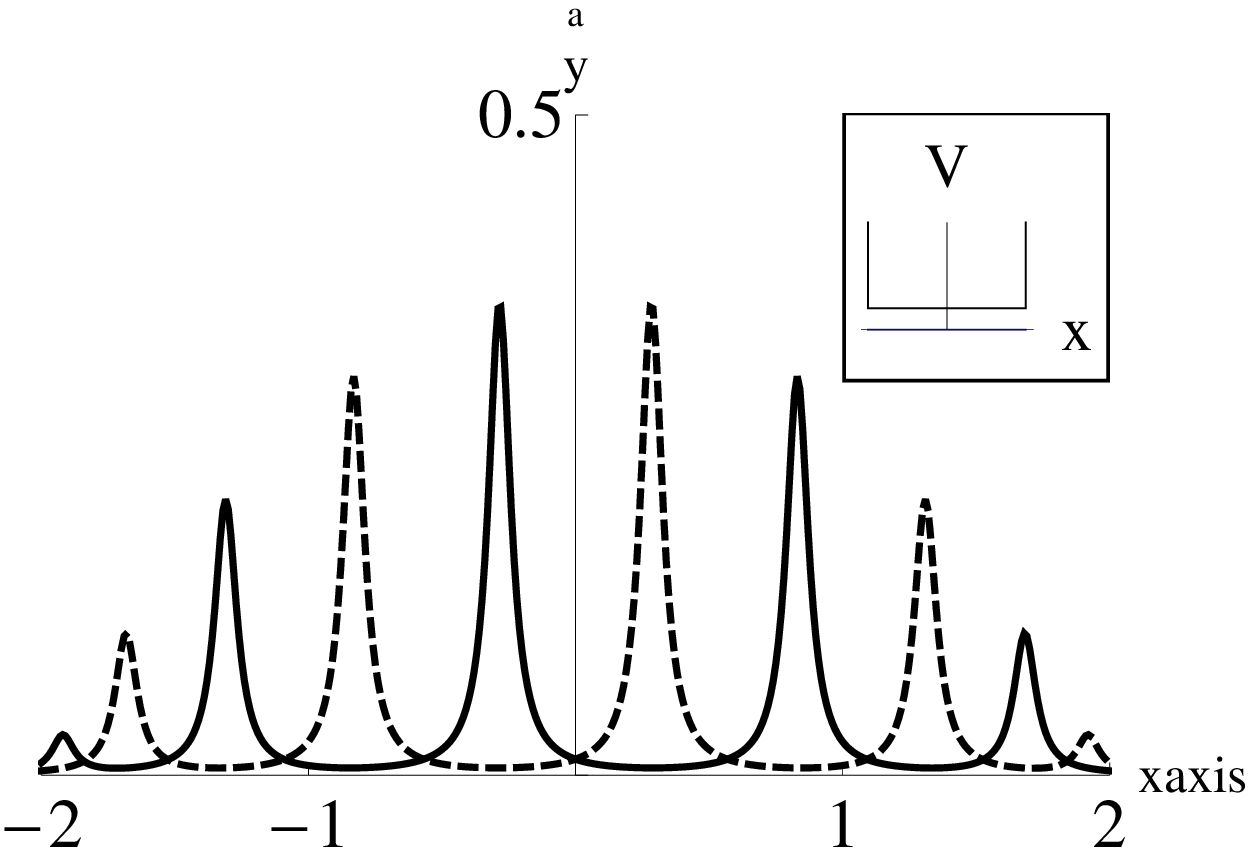}\includegraphics[width=0.5\columnwidth]{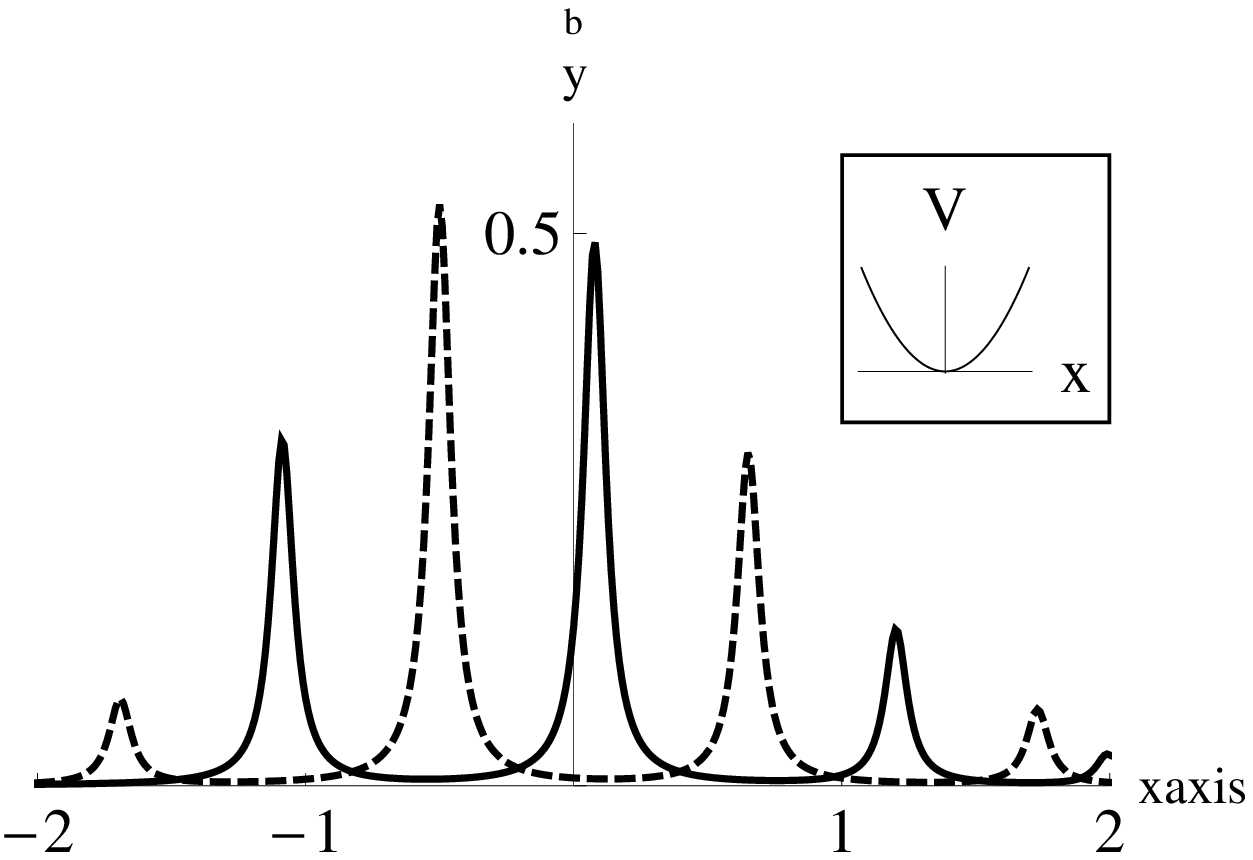}
   \caption{The electromagnetic absorption spectrum when atoms are excited from the ``a'' state to the ``c'' state with open boundary conditions. (a) The atoms are trapped in a lattice with 10 sites and hard walls at its ends. (b) The atoms are trapped in a harmonic trap so that the distance between the Majorana modes is 10 sites. In (b), $\mu_c$ refers to the chemical potential of the ``c'' atoms at the center of the trap. The solid curve is the absorption spectrum of $\gplus$, and the dashed curve of $\gminus$. Insets show the confining potential minus the optical lattice.}
 \label{fig:GammaOpen}
\end{figure}

Here we consider an experimentally simpler case where both the ``a'' and the ``c'' atoms are trapped in a linear geometry by an optical lattice of length $L$ and lattice spacing $a$ and with hard-wall boundary conditions. The wave functions of the ``c'' atoms are
\begin{equation}
 \psi_k(j) = \frac{\sqrt{2}\sin kr_j}{\sqrt{N}},
\end{equation}
where the allowed values of $k$ are $n\pi/(N+1)$ for integer values of $n$. The spectra, given by
\begin{multline}
 \Gamma(\gpm,\omega) = \frac{4\pi|\Omega|^2}{N\hbar}\sum_k\delta(\omega-\epsilon_k)\\ \times\left|\sum_jf_0(j)\left(\sin kr_j \pm \sin k((N+1)a-r_j)\right)\right|^2,
  \label{eqn:Gamma open}
\end{multline}
now have a richer, inter-digitated structure. They are plotted in Fig. \ref{fig:GammaOpen}(a) for a small lattice in the special case of Eq. (\ref{eqn:trivial case}), where $f_0(j)=\delta_{j1}$.

Figure \ref{fig:GammaOpen} illustrates that the spectra exhibit $N/2$ oscillations within a frequency range equal to the bandwidth of the ``c'' atoms. As the system becomes longer, the spectra develop more closely spaced oscillations. In the thermodynamic limit, high-frequency resolution is needed to differentiate $\gplus$ and $\gminus$. Depending on the bandwidth of the ``c'' atoms and the experimentally achievable frequency resolution, this places an upper limit on the length of the superfluid. A lower limit is placed by the hybridization of the edge modes. In experiments involving 1D optical lattices \cite{Hulet2010}, $N$ is typically of the order of 100. As explained at the beginning of this section, the bandwidth of the ``c'' atoms must be less than the superfluid gap in order to separate the spectral features of the bulk and edge modes. As explained in Sec. \ref{sec:Exptal Setup}, the superfluid gap produced by different techniques is in the kHz--MHz range. Therefore, the frequency resolution needed to resolve the interdigitated absorption spectra is on the order of kHz.

\subsection{Harmonic Traps}\label{subsec:harmonic}
Finally we consider the most experimentally realistic case where both the ``a'' atoms and ``c'' atoms are trapped by a harmonic potential. The effect of introducing a harmonic trap is twofold: The individual peaks shift to higher frequencies while the envelope shifts to lower frequencies [see Fig. \ref{fig:GammaOpen}(b)]. Qualitatively, however, the spectrum is nearly indistinguishable from a hard-wall trap. We numerically calculate and plot the result for a small lattice in a harmonic trap in Fig. \ref{fig:GammaOpen}(b).

\section{Dynamics} \label{sec:Dynamics}
To use our system to process quantum information we need algorithms to implement quantum gates. Here we construct gates based on microwave illumination. Each photon absorbed or emitted flips an atom between the ``a'' state and the ``c'' state, creating a quasiparticle and changing the particle number parity of the ``a'' atoms. In the appropriate frequency range, this parity change corresponds to flipping between $\gplus$ and $\gminus$. We envision that we can utilize this feature to perform gate operations. For example, if a microwave pulse is applied to a qubit in a state $|\psi_i\rangle = \alpha\gminus\otimes|0\rangle + \beta\gplus\otimes|0\rangle$, the qubit will evolve to $|\psi_f\rangle = \alpha(\cos\phi_-\gminus\otimes|0\rangle +\sin\phi_-\gplus\otimes|c_1\rangle) + \beta(\cos\phi_+\gplus\otimes|0\rangle+\sin\phi_+\gminus\otimes|c_2\rangle)$ after a time $t$, where $|0\rangle$ is the vacuum of ``c'' atoms, $|c_1\rangle$ and $|c_2\rangle$ are single-particle ``c'' states, and $\otimes$ is the Cartesian product.

In the strongly coupled regime $\Omega>J_c$ and the weakly coupled regime $\Omega<J_c/N$, we show in Sec. \ref{subsec:gates fast} that we can arrange $\phi_-=\pm\pi/2, \phi_+=\pm\pi/2$ and $|c_1\rangle=|c_2\rangle$. Under these circumstances, the ``c'' atoms are disentangled from the qubit, and we have produced an $X$ or a $Y$ gate (so named because they act as Pauli operators $\sigma_x$ or $\sigma_y$). The implementation of the $X$ and $Y$ gates will be explicitly demonstrated in Sec. \ref{subsec:gates fast}. Composing these operations yields a $Z$ gate. We show in Sec. \ref{sec:ancillary bit} that more complicated (universal) gates can be implemented by using composite logical qubits.

In the intermediate regime $J_c>\Omega>J_c/N$, we show in Sec. \ref{subsec:gates intermediate} that can arrange $\phi_-=\pi/2, \phi_+=0$. Here the final state will be $|\psi_f\rangle = \gplus\otimes(\alpha|c_1\rangle+\beta|0\rangle)$. The ``c'' atoms are in a superposition of number eigenstates, but because of the product structure, they have no impact on future gate operations. Thus, this acts as a SET gate which projectively initializes the qubit in $\gplus$. To set the qubit in $\gminus$, we arrange $\phi_+=\pi/2, \phi_-=0$.

Since the gates substantially change the state, they cannot be described by linear response. This motivates us to study the full quantum dynamics in Sec. \ref{sec:numerical methods}. In Sec. \ref{sec:gates} we propose algorithms to implement various gates, and use the numerical methods described in Sec. \ref{sec:numerical methods} to observe the dynamics during the gate operations.

\section{Numerical Methods} \label{sec:numerical methods}
To study the dynamics of the system we work in the Heisenberg picture. Within a mean-field approximation, the Heisenberg equations of motion,
\begin{equation}\begin{array}{rlc}
i\partial_t\ha_j(t) &=& [\ha_j(t),\hat{H}],\\  i\partial_t\hc_j(t) &=& [\hc_j(t),\hat{H}], \label{eqn:Heisenberg}
\end{array}\end{equation}
reduce to a matrix equation $i\partial_tX(t)=H(t)X(t)$, where $H(t)$ is a $4N\times4N$ matrix, and $X(t)$ is a $4N\times1$ vector that maps fermionic operators at time $t$ to those at $t=0$,
\begin{multline}
\left(\begin{array}{cccccccccc}
\ha_1(t) & \ha_2(t) & .. & \ha_N(t)  \ha\+_1(t) & .. & \ha_N\+(t) & \hc_1(t) & .. & \hc_N\+(t) \end{array}\right)^T\\
= X(t) \left(\begin{array}{cccccccccc}\ha_1(0)& .. & \ha_N\+(0) & \hc_1(0) & .. & \hc_N\+(0) \end{array}\right)^T.
\end{multline}
In a typical numerical experiment we take $N\simeq50$. The matrix $H(t)$ can be computed from Eq. (\ref{eqn:Heisenberg}). We update $X$ via $X(t+\delta t)=e^{-i \overline{H}\delta t}X(t)$, where $\overline{H}$ is an approximant to the average $H(t)$ in the interval between $t$ and $t+\delta t$. In the coupled-wire case, $H(t)$ depends on $X(t)$ through the self-consistency condition in Eq. (\ref{eqn:self-consistency}). This self-consistent approach conserves total mean particle number of ``a'' and ``c'' atoms, and is a more sophisticated generalization of the random phase approximation. In the proximity-induced superfluid case, contact with a bath implies $H(t)$ is a constant and the total mean number of ``a'' and ``c'' atoms are not conserved.

From $X(t)$, we calculate $\langle\hmaj\+\hmaj\rangle$, the probability of the qubit being in the state $\gplus$. We also calculate coherences between $\gplus$ and $\gminus$ through $\langle\hmaj\rangle$. We quantify the success of our protocols through the fidelity of the final state produced by the gates, defined as the norm of the overlap of the final state $|\psi_f\rangle$ with the intended target state $|\phi\rangle$, $f(|\phi\rangle,|\psi_f\rangle) = ||\langle\phi|\psi_f\rangle||$. We express the fidelity in terms of expectation values of edge-state creation and annihilation operators. For example if the intended target state is $\gplus$, the fidelity is $\sqrt{\langle\psi_f|\hmaj\+\hmaj|\psi_f\rangle}$. With these tools we explore algorithms to implement various gates on qubits.

\section{Quantum Gates} \label{sec:gates}

\subsection{Intermediate Regime: Projective Initialization} \label{subsec:gates intermediate}
We begin by discussing the physics of the intermediate regime $J_c>\Omega>J_c/N$ as this is the most familiar. Here the photon absorption rate is given by Fermi's golden rule, and Figs. \ref{fig:GammaPeriodic} and \ref{fig:GammaOpen} can be directly interpreted as the rates of producing ``c'' atoms. To perform a projective initialization, we use the fact that there are frequencies $\omega$ where $\Gamma(\gplus,\omega)\neq0$, but $\Gamma(\gminus,\omega)=0$. Logically one must be able to deterministically set the qubit into the state $\gminus$ by shining photons at those frequencies. A similar method can be used to set the qubit into $\gplus$. By working in the regime $J_c/N<\Omega<J_c$, we ensure that any atoms which exit in the ``c'' state carry away the information about the initial state of the qubit.

\begin{figure}[htbp]
   \unitlength=1in
   \psfragscanon
   \psfrag{x}[][][1.1]{\begin{picture}(0,0)
    \put(-0.3,-0.1){\makebox(0,0)[l]{
    $\frac{\Omega t}{\hbar}$
   }}\end{picture}}
   \psfrag{y}[][]{\begin{picture}(0,0)
    \put(0.1,0.05){\makebox(0,0){
    $\langle\hmaj\+\hmaj\rangle$
   }}\end{picture}}
   \psfrag{a}[][]{\begin{picture}(0,0) \put(0,-1.3){\makebox(0,0){(a)}} \end{picture}}
   \psfrag{b}[][]{\begin{picture}(0,0) \put(0,-1.3){\makebox(0,0){(b)}} \end{picture}}
   \psfragscanoff
   \includegraphics[width=0.5\columnwidth]{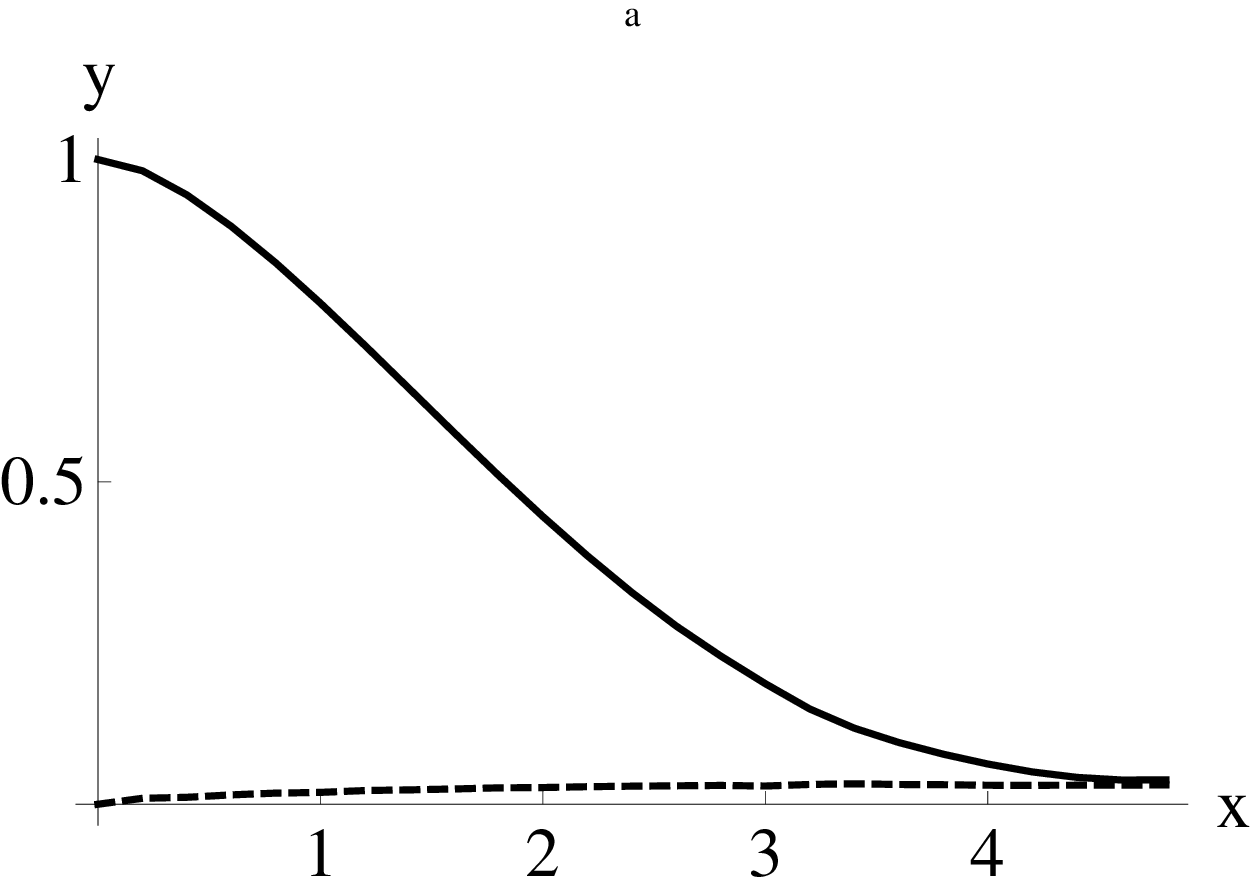}\includegraphics[width=0.5\columnwidth]{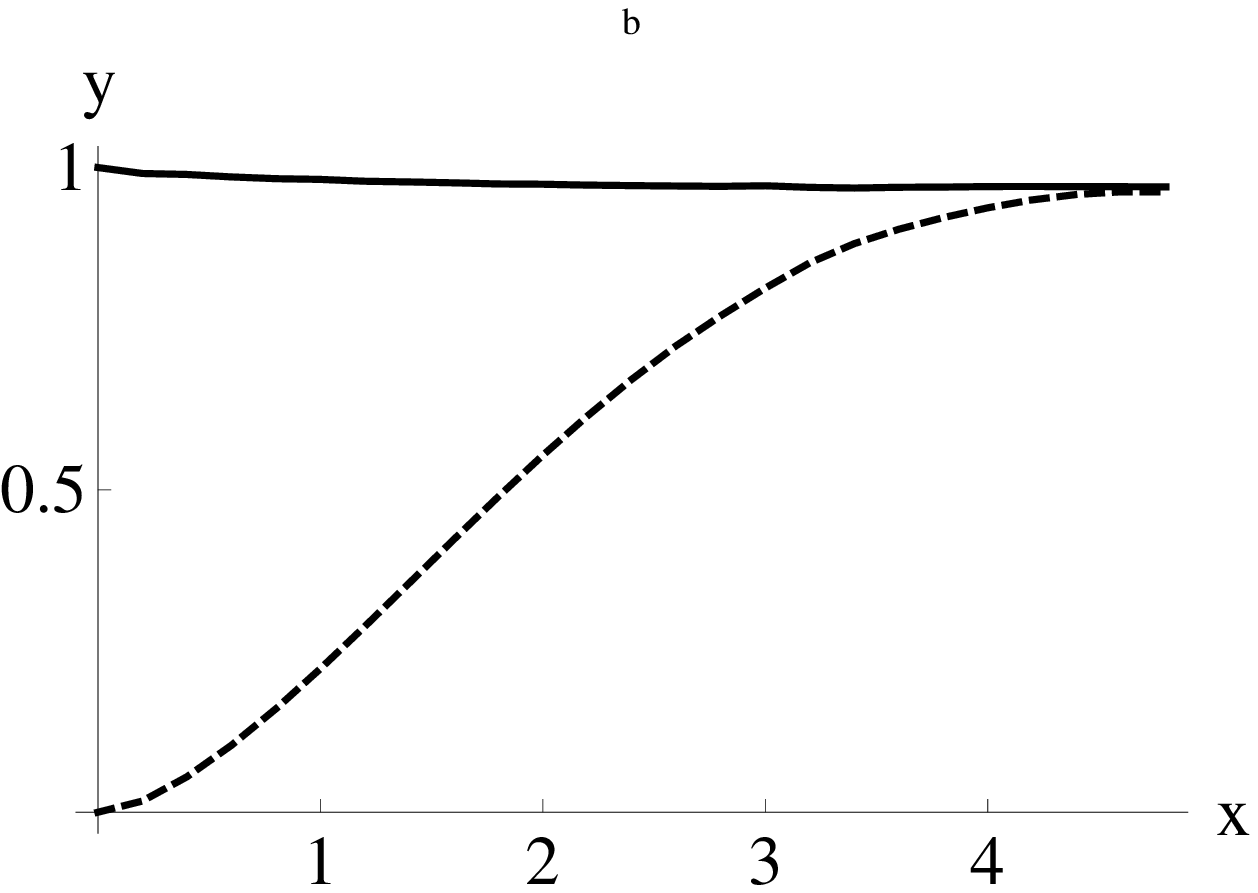}
   \caption{Dynamics of $\langle\hmaj\+\hmaj\rangle$ during the initialization gate in a geometry where the system consists of a single wire immersed in a superfluid cloud. The wire is in the form of a ring, as explained in Sec. \ref{subsec:periodic}. Electromagnetic waves are shined upon the system, with frequencies (a) $\omega = -\mu_c-2J_c$ and (b) $\omega = -\mu_c+2J_c$. The solid curve corresponds to the system's initial state being $|\psi\rangle = \gplus$, and the dashed curve to $|\psi\rangle = \gminus$. The fidelities of the initialization gate in the two cases are 
   nearly $100\%$.}
   \label{fig:successful SET}
\end{figure}

Figure \ref{fig:successful SET} shows the dynamics of $\langle\hmaj\+\hmaj\rangle$ for the proximity-induced case (i.e, when $\Delta_j$ are constant) for two different microwave frequencies and two different initial states $\gplus$ and $\gminus$. In Fig. \ref{fig:successful SET}(a) we choose a frequency where $\Gamma(\gminus,\omega)=0$, producing $\gminus$ as the final state for both initial states, and in Fig. \ref{fig:successful SET}(b) we choose a frequency where $\Gamma(\gplus,\omega)=0$, producing $\gplus$ as the final state for both initial states. The fidelity of the final state produced in all four cases is nearly $100\%$.

\begin{figure}[htbp]
   \unitlength=1in
   \psfragscanon
   \psfrag{x}[][][1.1]{\begin{picture}(0,0)
    \put(-0.35,-0.15){\makebox(0,0)[l]{
    $\frac{\Omega t}{\hbar}$
   }}\end{picture}}
   \psfrag{y}[][]{\begin{picture}(0,0)
    \put(0.1,0.05){\makebox(0,0){
    $\langle\hmaj\hmaj\+\rangle$
   }}\end{picture}}
   \psfrag{a}[][]{\begin{picture}(0,0) \put(0,-1.3){\makebox(0,0){(a)}} \end{picture}}
   \psfrag{b}[][]{\begin{picture}(0,0) \put(0,-1.3){\makebox(0,0){(b)}} \end{picture}}
   \psfragscanoff
   \includegraphics[width=0.5\columnwidth]{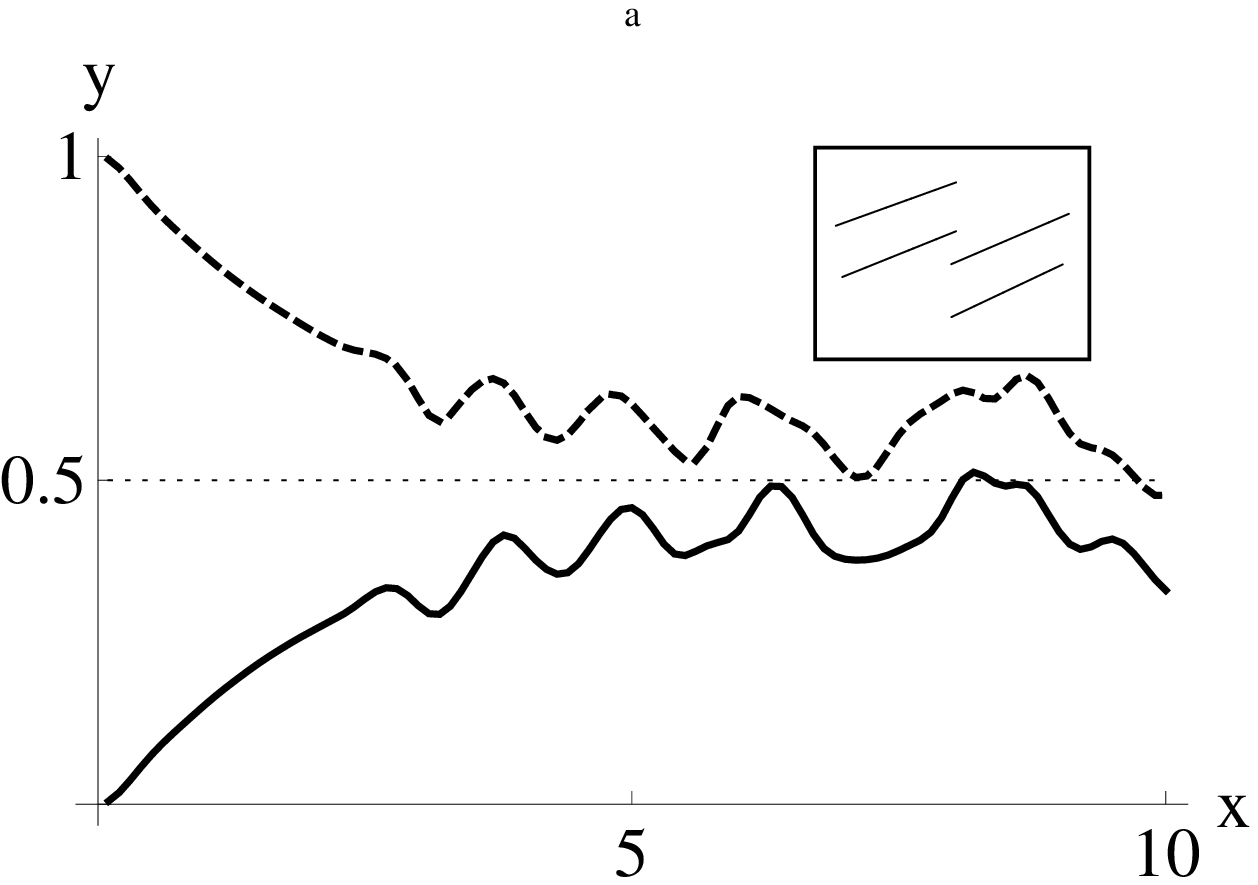}\includegraphics[width=0.5\columnwidth]{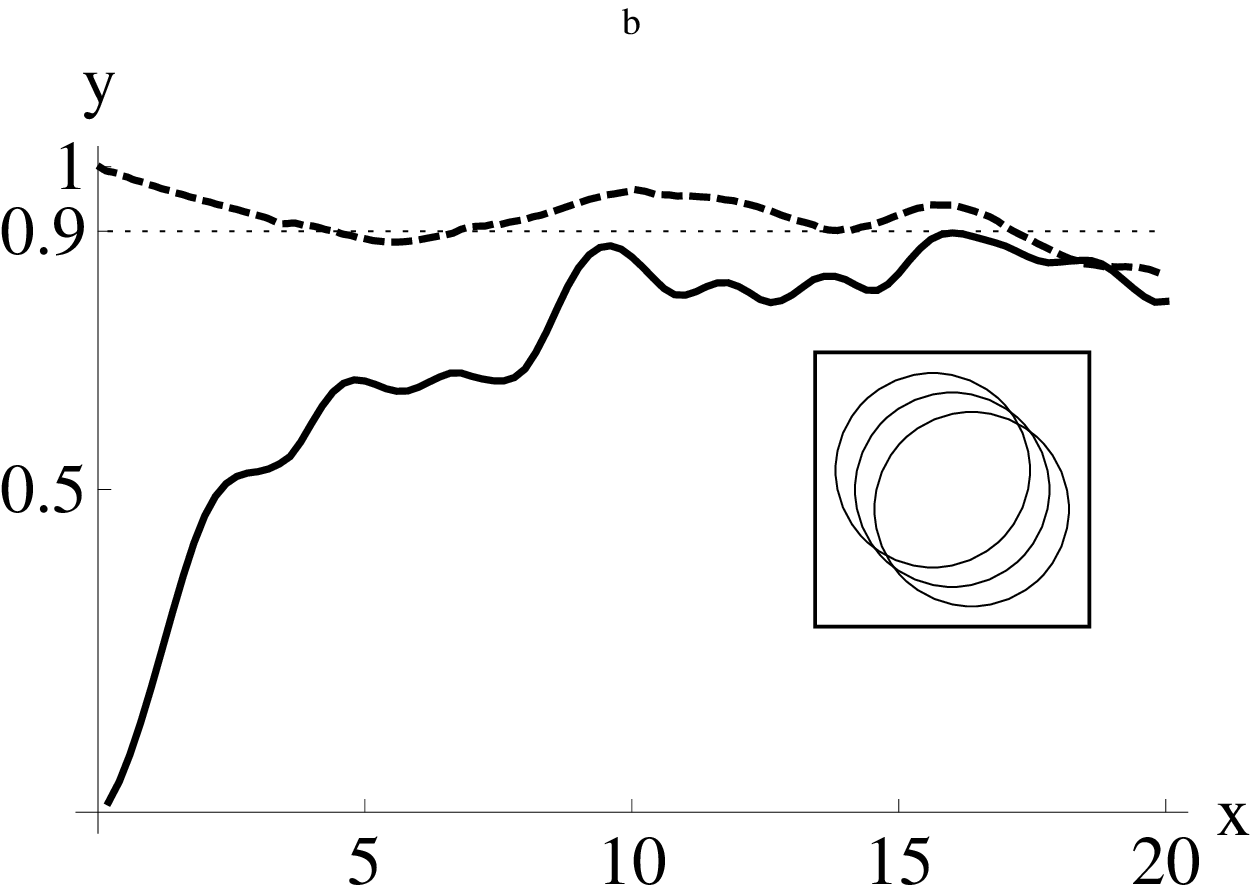}
   \caption{Dynamics of $\langle\hmaj\hmaj\+\rangle$ during the initialization gate in a geometry of coupled wires or rings. (a) The atoms are trapped in an array of linear wires. Parameters: $\Delta_j=J=40\Omega, J_c=5\Omega, N=25, \omega_0=-\mu_c$. (b) The atoms are trapped in an array of rings. Parameters: $\Delta_j=J=20\Omega, J_c=7.5\Omega, N=50, \omega_0=-\mu_c-2J_c$. Insets show the geometry of trapping. The solid line represents the case where the system is initially in the state $\gplus$, and the dashed line represents where it is initially in $\gminus$. A frequency $\omega_0$ where $\Gamma(\gminus,\omega_0)=0$ is shined upon the system, with the intention of setting the qubit to $\gminus$. The square of the fidelity of the final state is roughly $50\%$ in (a) and $90\%$ in (b). In both cases, there is some absorption when the initial state is $\gminus$ due to finite bandwidth of the ``c'' atoms. In both cases, this shifts $\omega$ to a new frequency where $\gminus$ is not dark. In case (a) the absorption at this new frequency is significant, while in case (b) the absorption is still small.}
   \label{fig:failed SET}
\end{figure}

We find that our algorithm to initialize the qubit does not work in the coupled-wire geometry, where $\Delta_j$ are determined self-consistently. Figure \ref{fig:failed SET} shows the dynamics of $\langle\hmaj\+\hmaj\rangle$ for this case. This failure is due to an induced chemical potential in the wires during photon absorption. We examine this failure in more detail below.

\subsubsection{Failure of Projective Initialization in Arrays of Coupled Wires}
As previously introduced, we consider an array of wires or rings, each initially in the state $\gplus$. We illuminate the atoms with photons of frequency $\omega_0$ where $\Gamma(\gplus,\omega_0)\neq0$, but $\Gamma(\gminus,\omega_0)=0$, with the intention of driving all the clouds into $\gminus$. If this process was successful, it would remove one ``a'' atom from each wire and hence shift the chemical potential. This resulting shift $\delta\mu$ would be of order $J/N$, where $J$ is the hopping amplitude of the ``a'' atoms and $N$ is the number of lattice sites. As detailed in Eq. (\ref{eqn:frequency relation}), the chemical potential enters into the relationship between the physical electromagnetic frequency $\omega_{\rm{physical}}$ and the frequency $\omega$ that appears in Fermi's golden rule. The condition $\Gamma(\gminus,\omega)=0$ will be violated at the new frequency $\omega=\omega_0-\delta\mu/\hbar$, and the gate instead drives the system into a steady-state mixture of $\gplus$ and $\gminus$. As a technical point, we do not have a time-dependent $\mu$ in our equations; rather this effect is manifest by a time-dependent phase for $\Delta_j$. Figure \ref{fig:failed SET}(a) shows a self-consistent mean-field theory calculation of the evolution of $\langle\hmaj\+\hmaj\rangle$ when the atoms are trapped in an array of finite wires. It illustrates that the square of the fidelity of the final state is only about $50\%$.

One might argue that the scaling of $\delta\mu\simeq O(J/N)$ with $N$ implies that the gate will work in the thermodynamic limit. Such an argument is fallacious when the atoms are trapped in a linear geometry. In this case the spacing $\delta\omega$ of the peaks in the absorption spectra scales as $O(J_c/N\hbar)$. In order to have localized edge modes where the edge spectrum is separated from the bulk, one needs $J_c<<\Delta\simeq J$. Therefore in this geometry, $\delta\mu$ is always large compared to $\hbar\delta\omega$, and the absorption rate in the state $\gminus$ at the new frequency $\omega=\omega_0-\delta\mu/\hbar$ is significant. The fidelity is therefore poor.

\begin{figure}[htbp]
   \unitlength=1in
   \psfragscanon
   \psfrag{norm}[][]{\begin{picture}(0,0)
    \put(0,0.05){\makebox(0,0){
    $||\langle\phi|\psi_f\rangle||$
   }}\end{picture}}
   \psfragscanoff
   \includegraphics[width=0.55\columnwidth]{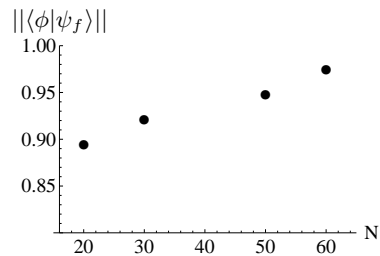}
   \caption{Fidelity of the initialization gate for different lengths of superfluid rings. The ``a'' and ``c'' atoms are confined in a ring geometry. We used parameters: $\Delta_j=J=20\Omega, J_c=7.5\Omega$, and $\omega_0=\mu_c\pm2J_c$ for desired final states $|\phi\rangle=|g_\pm\rangle$. $|\psi_f\rangle$ denotes the actual final state produced. The initialization gate has higher fidelity for larger lengths $N$.}
   \label{fig:fidelity}
\end{figure}

Conversely, the initialization gate works in the thermodynamic limit for an array of rings. As shown in Fig. \ref{fig:GammaPeriodic} for the ring geometry, the change in the spectral weights over a scale of $O(J_c/N)$ is small. Therefore, in this case one can spectroscopically set the qubit with higher fidelity. This success is illustrated in Fig. \ref{fig:failed SET}(b) for parameters where the square of the fidelity of the final state is roughly $90\%$. Figure \ref{fig:fidelity} illustrates that the fidelity grows as $N$ increases, albeit slowly.

We avoid these difficulties in the rest of this paper by restricting ourselves to the case of a single wire in proximity to a superfluid cloud. Similar physics can be seen in arrays of coupled wires, but the changing chemical potential will result in smaller fidelities.

\subsection{Fast and Slow Regimes: Coherent Gate Operations} \label{subsec:gates fast}
In this section we explore algorithms to perform a set of gates required for quantum computing: $X, Y, H$ (Hadamard) and $C^Y$ (controlled-$Y$). It will be convenient to also define a $Z$ gate: $Z=-iXY$. The $X, Y, Z$, and $H$ gates act on single qubits and are described below. The $C^Y$ gate is a two-qubit gate which will be analyzed in Sec. \ref{sec:ancillary bit}.

The $X$, $Y$ and $Z$ gates perform a rotation of the Bloch sphere by $\pi$ radians around the $x,\ y$ and $z$ axes. That is,
\begin{equation}\begin{array}{rlcrlc}
X\gplus &=& \gminus, X\gminus &=& \gplus, \\
Y\gplus &=& i\gminus, Y\gminus &=& -i\gplus, \\
Z\gplus &=& \gplus, Z\gminus &=& -\gminus.
\end{array}\end{equation}
For our system, these gates can be expressed as $X=\hmaj+\hmaj\+,\ Y=i(\hmaj-\hmaj\+)$, and $Z=\hmaj\+\hmaj-\hmaj\hmaj\+$. The $X$ and $Y$ gates are related to each other by a gauge transformation of the fermionic creation and annihilation operators. To disambiguate the situation we take $\Delta_j>0$. It can be observed from Eq. (\ref{eqn:f0 solution}) that, in this case, the coherence factors $f_0(j)$ are peaked at $j=1$.

The $H$ gate is
\begin{equation}\begin{array}{rlc}
H\gplus &=& \frac{\gplus+\gminus}{\sqrt{2}}, \\
H\gminus &=& \frac{\gplus-\gminus}{\sqrt{2}}.
\end{array}\end{equation}
It can be expressed in terms of edge mode operators as $H = (\hmaj+\hmaj\+ + \hmaj\+\hmaj-\hmaj\hmaj\+)/\sqrt{2}$. The $H$ gate creates superpositions between states of opposite number parity. Since the Hamiltonian modeling our system [Eq. (\ref{eqn:Hfull})] is parity-conserving, it is not possible to implement the $H$ gate using microwaves on these qubits: Any attempt to produce it using microwaves will result in a state with entanglement between the ``a'' and ``c'' atoms. We explain how to avoid this in Sec. \ref{sec:ancillary bit}, where we propose an architecture in which two logical qubits are encoded in three physical qubits. The logical gates will be made from $X, Y$, and $Z$ rotations on physical qubits, which we explore below.

\subsubsection{$X$ Gate} \label{subsubsec:X}
To implement the $X$ gate on a physical qubit, we illuminate only the left half of the qubit. Experimentally, this can be done by either placing a mask on the right half or by focusing light of short wavelength on the left half of the system, i.e, using Raman techniques with light of wavelength shorter than half the size of the system $\simeq$ O(mm). In the rotating-wave approximation, the term in the Hamiltonian that involves electromagnetic waves [Eq. (\ref{eqn:HMW})] is now
\begin{equation}
 \hat{H}_X = \Omega\sum_{j\leq N/2} \hc_j\+ \ha_j + h.c..
 \label{eqn:HX}
\end{equation}
We can decompose $\hat{H}_X$ as $\hat{H}_X = \hat{H}_X^{\rm{res}} + \hat{H}_X^{\rm{off-res}}$, where $\hat{H}_X^{\rm{res}}$ includes terms in which edge modes are resonant with the ``c'' states, and $\hat{H}_X^{\rm{off-res}}$ includes all the other terms which are off-resonant. This can be observed by rewriting $\ha_j$ in terms of quasiparticle operators. Inverting Eq. (\ref{eqn:u,v}) and using the definition of $f_0(j)$, we obtain
\begin{equation}\begin{array}{cl}
 \ha_j = & f_0(j)(\hmaj+\hmaj\+) + f_0(N+1-j)(\hmaj-\hmaj\+) \\
 &+ \sum_{\nu\neq0} (u_\nu(j)\hat{\gamma}_\nu + v_\nu(j)\hat{\gamma}_\nu\+).
\end{array}\end{equation}
The terms involving bulk quasiparticles ($\nu\neq0$) are off-resonant and are included in $\hat{H}_X^{\rm{off-res}}$. Further in Eq. (\ref{eqn:HX}), $\hc_j\+$ can be decomposed into ``c'' eigenmode creation operators. All the eigenmodes will be excited in the fast regime $\Delta_j>>\Omega>J_c$, while only one eigenmode will be resonantly coupled in the slow regime $\Omega_c<J_c/N$. Terms in Eq. (\ref{eqn:HX}) involving modes that are not excited are included in $\hat{H}_X^{\rm{off-res}}$. Hereafter, we neglect the off-resonant terms, $\hat{H}_X^{\rm{off-res}}$. We explicitly consider the form of $\hat{H}_X^{\rm{res}}$ in both the slow ($\Omega<J_c/N$) and the fast ($\Omega>J_c$) limits.

\begin{figure}[htbp]
   \unitlength=1in
   \psfragscanon
   \psfrag{x}[][][1.1]{\begin{picture}(0,0)
    \put(-0.35,-0.15){\makebox(0,0)[l]{
    $\frac{\Omega t}{\hbar}$
   }}\end{picture}}
   \psfrag{x2}[][][1.1]{\begin{picture}(0,0)
    \put(-0.35,0.1){\makebox(0,0)[l]{
    $\frac{\Omega t}{\hbar}$
   }}\end{picture}}
   \psfrag{prob}[][]{\begin{picture}(0,0)
    \put(0.1,0.05){\makebox(0,0){
    $\langle\hmaj\+\hmaj\rangle$
   }}\end{picture}}
   \psfrag{coherence}[][]{\begin{picture}(0,0)
    \put(0.1,0.05){\makebox(0,0){
    $\langle\hmaj\rangle$
   }}\end{picture}}
   \psfrag{a}[][]{\begin{picture}(0,0) \put(0,-1.36){\makebox(0,0){(a)}} \end{picture}}
   \psfrag{b}[][]{\begin{picture}(0,0) \put(0,-1.21){\makebox(0,0){(b)}} \end{picture}}
   \psfrag{c}[][]{\begin{picture}(0,0) \put(0,-1.36){\makebox(0,0){(c)}} \end{picture}}
   \psfrag{d}[][]{\begin{picture}(0,0) \put(0,-1.16){\makebox(0,0){(d)}} \end{picture}}
   \psfragscanoff
  \includegraphics[width=0.5\columnwidth]{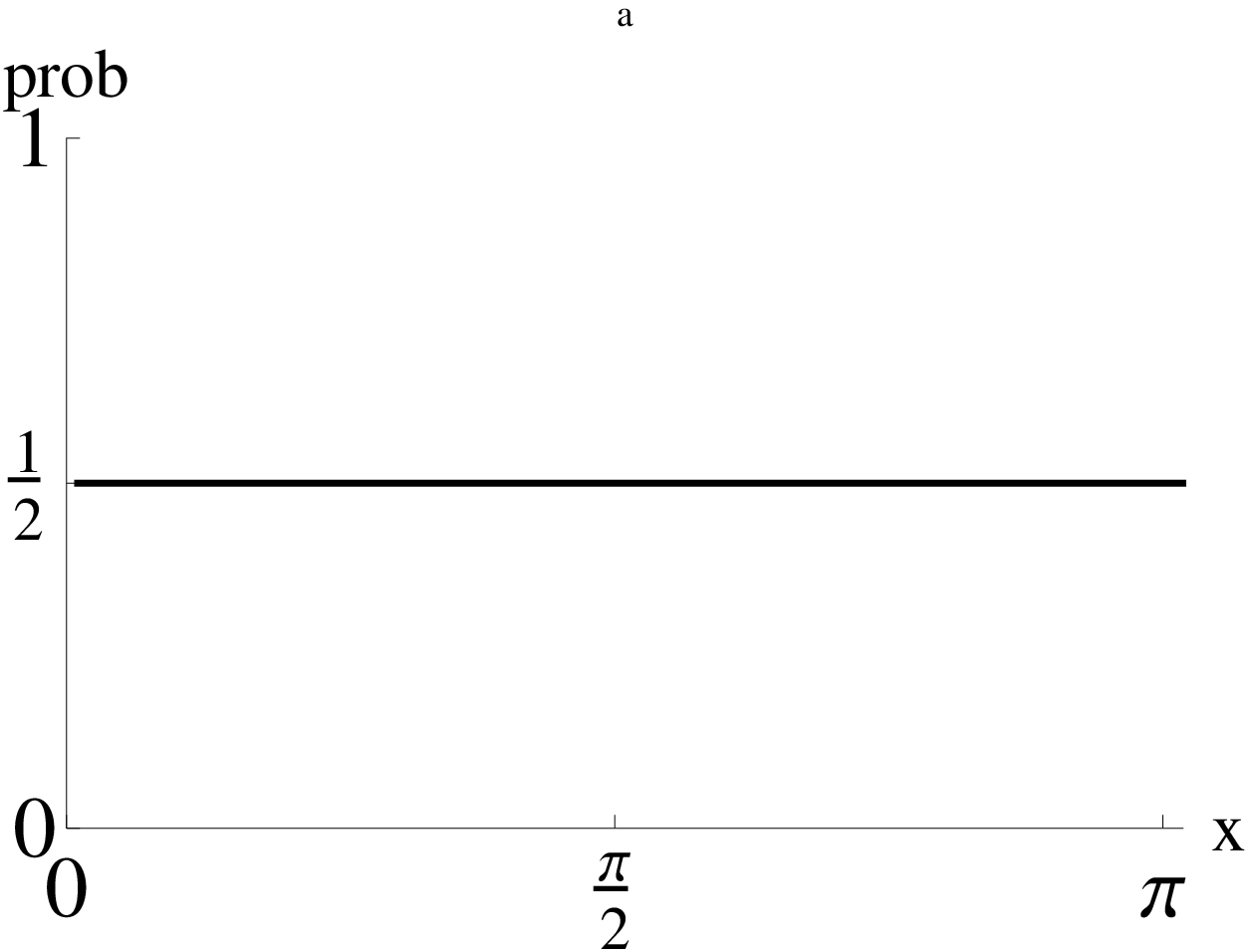}\includegraphics[width=0.5\columnwidth]{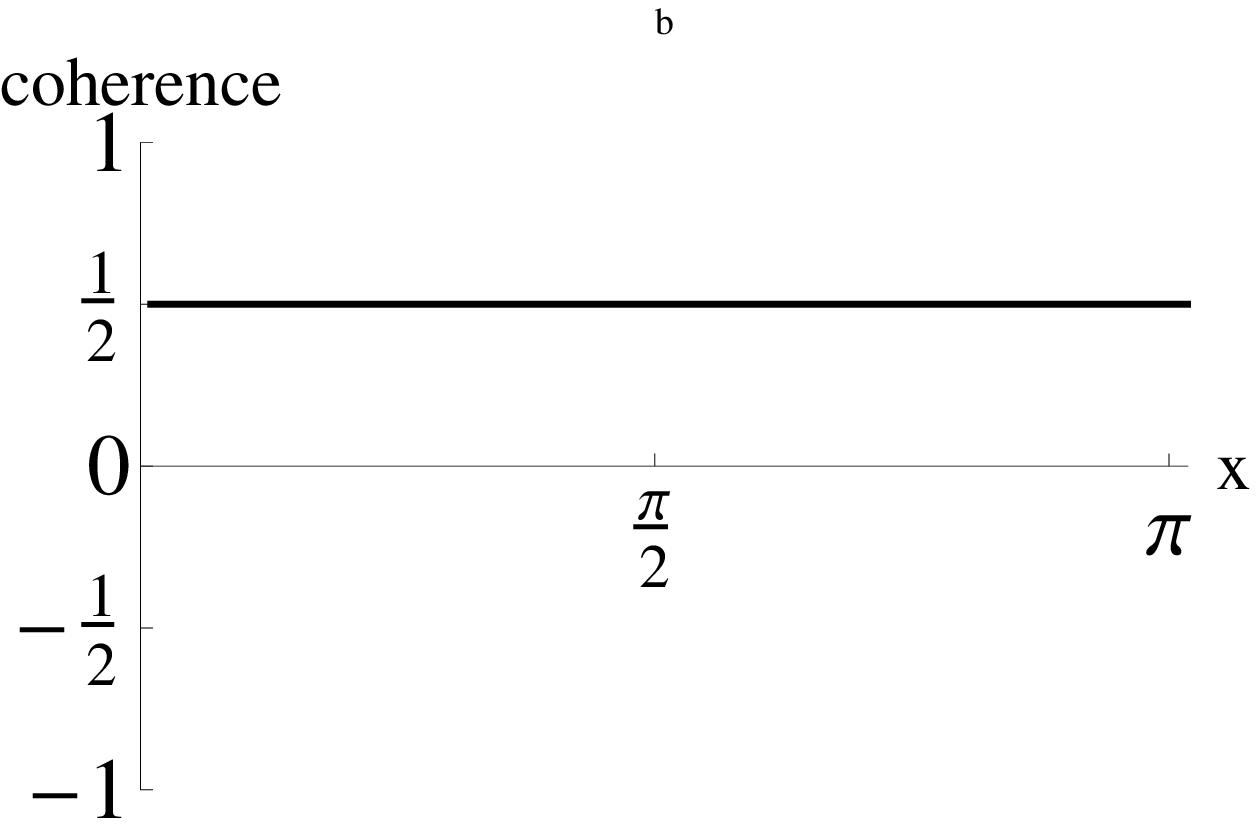}
  \\[0.2in]
  \includegraphics[width=0.5\columnwidth]{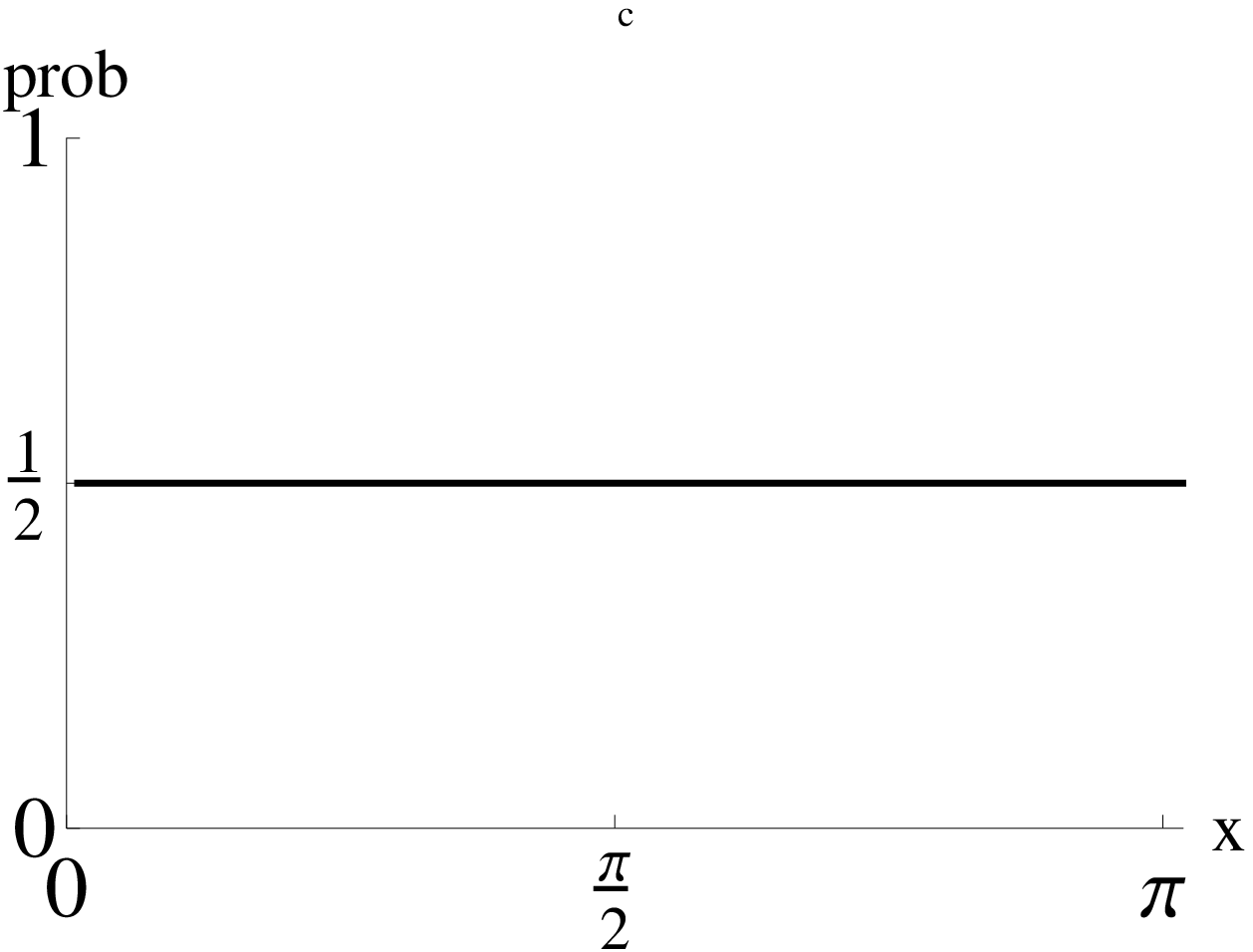}\includegraphics[width=0.5\columnwidth]{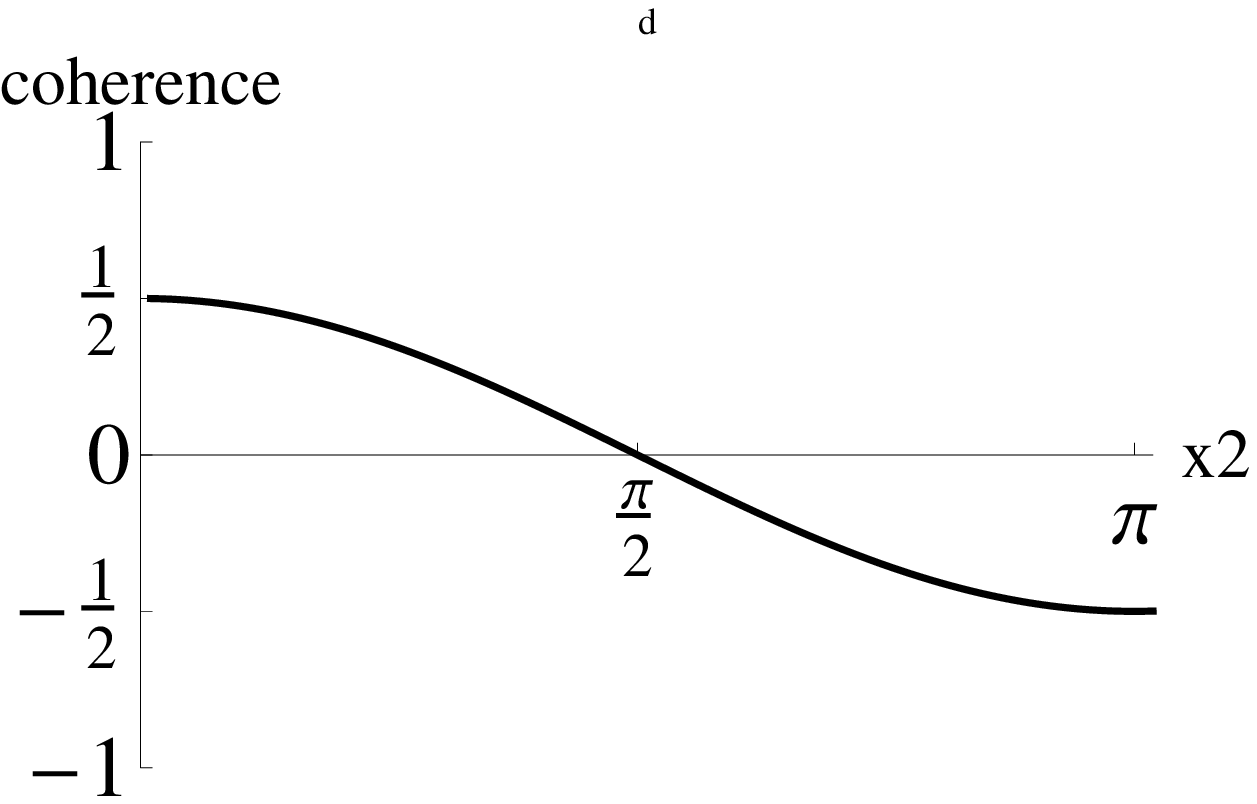}
   \caption{Dynamics of $\langle\hmaj\+\hmaj\rangle$ and $\langle\hmaj\rangle$ during the $X$ and $Y$ gates. The parameters used are $\Delta=J=80\Omega, \mu=0, J_c=0, \omega=-\mu_c$, and the system is initialized to $\frac{1}{\sqrt{2}}(\gplus+\gminus)$. Panels (a) and (b) illustrate the dynamics in the case of an $X$ gate operation and show that the final state is $\frac{1}{\sqrt{2}}(\gplus+\gminus)$. Panels (c) and (d) illustrate the dynamics in the case of a $Y$ gate operation and show that the final state is $\frac{i}{\sqrt{2}}(\gplus-\gminus)$.}
   \label{fig:gates}
\end{figure}

As explained above, in the fast regime $\Delta_j>>\Omega>J_c$,
\begin{equation}
 \hat{H}_X^{\rm{res}} = \frac{\Omega}{2}\sum_{j\leq N/2}(f_0(j)\hc_j\+X - \\ i f_0(N+1-j)\hc_j\+Y + h.c),
\end{equation}
where $X=\hmaj+\hmaj\+, Y=i(\hmaj-\hmaj\+)$, and $f_0(j)$ are the coherence factors discussed in Sec. \ref{sec:Majorana modes}. Since $f_0(j)$ falls exponentially to zero away from $j=1$, the coefficients of $Y$ are small and can be ignored. We define $\hat{\overline{c}}\+ = \frac{\sum_{j\leq N/2} f_0(j)\hc_j\+}{\sum_{j\leq N/2}|f_0(j)|^2}$ and $\tilde{\Omega} = \Omega\sum_{j\leq N/2}|f_0(j)|^2$ to produce
\begin{equation}
\hat{H}_X^{\rm{res}} = \frac{\tilde{\Omega}}{2}(\hat{\overline{c}}\+-\hat{\overline{c}})X.
\end{equation}
For short times $t<<\hbar/J_c$, one can ignore the dynamics of ``c'' atoms. The state of the qubit at time $t$ is
\begin{equation}
|\psi(t)\rangle = \left(\cos\frac{\tilde\Omega t}{2\hbar}+\sin\frac{\tilde\Omega t}{2\hbar}\frac{\hat{\overline{c}}-\hat{\overline{c}}\+}{i}X\right)|\psi(0)\rangle + O(J_ct/\hbar).
\end{equation}
The $X$ gate is implemented by shining a $\pi$ pulse lasting $T=\frac{\pi\hbar}{\tilde\Omega}$. The pulse also excites or deexcites a ``c'' atom, but, as explained in Sec. \ref{sec:Dynamics}, these ``c'' atoms do not play any role in future dynamics. The dynamics of $\langle\hmaj\+\hmaj\rangle$ and $\langle\hmaj\rangle$ for a $\pi$ pulse are illustrated in Figs. \ref{fig:gates}(a) and \ref{fig:gates}(b) for an initial state $\frac{\gplus+\gminus}{\sqrt{2}}$, which is an eigenstate of $X$. The average occupation of the edge mode, $\langle\hmaj\+\hmaj\rangle$, does not change, indicating that the probability of the qubit being in $\gplus$ remains $50\%$ throughout the pulse. The coherence $\langle\hmaj\rangle$ also remains constant, indicating that the phase between $\gplus$ and $\gminus$ remains zero.

In the slow regime $\Omega<J_c/N$, the Majorana modes only resonantly couple to one eigenmode of the ``c'' state. This eigenmode is a momentum state in translationally invariant geometries. In this regime,
\begin{equation}
 \hat{H}_X^{\rm{res}} = \frac{\Omega}{2}\left((\alpha_k\hc_k\+-\alpha_k^*\hc_k)X + i(\beta_k\hc_k\++\beta_k^*\hc_k)Y\right).
\end{equation}
where $k$ labels the spectrally selected mode, $\alpha_k=\frac{1}{2}\sum_{j\leq N/2}f_0(j)\psi_k(j),\ \beta_k=\frac{1}{2}\sum_{j\leq N/2}f_0(N+1-j)\psi_k(j)$, and $\psi_k$ is the wave function of the ``c'' mode discussed in Sec. \ref{sec:Gamma}. As before we neglect the coefficients of $Y$ and arrive at a similar expression
\begin{equation}
|\psi(t)\rangle = \left(\cos\frac{|\alpha_k|\Omega t}{2\hbar}+\sin\frac{|\alpha_k|\Omega t}{2\hbar}\frac{\alpha_k^*\hc_k-\alpha_k\hc_k\+}{i}X\right)|\psi(0)\rangle.
\end{equation}
The $X$ gate is implemented by shining a $\pi$ pulse lasting $T=\frac{\pi\hbar}{|\alpha_k|\Omega}$.

\subsubsection{$Y$ Gate} \label{subsubsec:Y}
To implement the $Y$ gate we illuminate only the right half of the system. The term in the Hamiltonian that involves electromagnetic waves [Eq. (\ref{eqn:HMW})] is replaced with
\begin{equation}
 \hat{H}_Y = \Omega\sum_{j> N/2} \hc_j\+ \ha_j e^{-i\omega t} + h.c. \label{eqn:Hy}
\end{equation}
Neglecting off-resonant terms and exponentially small terms, Eq. (\ref{eqn:Hy}) reduces to
\begin{equation}
 \hat{H}_Y^{\rm{fast}} = \frac{i\tilde{\Omega}}{2}(\hat{\overline{c}}\+ + \hat{\overline{c}})Y
\end{equation}
in the fast regime, and to
\begin{equation}
 \hat{H}_Y^{\rm{slow}} = \frac{i\Omega}{2}(\beta_k^*\hc_k+\beta_k\hc_k\+)Y
\end{equation}
in the slow regime. Here $\tilde\Omega, \hat{\overline{c}}$ and $\beta_k$ have the same meaning as in Sec. \ref{subsubsec:X}. The $Y$ gate is implemented by shining a $\pi$ pulse lasting $T=\frac{\pi\hbar}{\tilde\Omega}$ in the fast regime and $T=\frac{\pi\hbar}{|\beta_k|\Omega}$ in the slow regime. The dynamics of $\langle\hmaj\+\hmaj\rangle$ and $\langle\hmaj\rangle$ for such a $\pi$ pulse in the fast regime are illustrated in Figs. \ref{fig:gates}(c) and \ref{fig:gates}(d). Here we see a $\pi$ phase introduced in the coherence between $\gplus$ and $\gminus$.

\subsection{$Z$ Gate}
The $Z=-i XY$ gate can be implemented by illuminating a $\pi$ pulse on the right half, followed by a $\pi$ pulse on the left half.

\section{Two Logical Qubits Composed Of Three Physical Qubits} \label{sec:ancillary bit}
As explained in Sec. \ref{sec:gates}, it is impossible to implement the $H$ gate on individual physical qubits as our approach cannot create superpositions of states with odd and even numbers of particles. This motivates us to consider a more sophisticated architecture where we construct $n$ logical qubits from $n+1$ physical qubits. Alternatively, one could encode each logical qubit in a pair of physical qubits (for example, \cite{DasSarmaQtmComputing}), but our encoding is more compact. In this section, we focus on the case $n=2$. We define our construction, and propose algorithms to produce all the quantum gates required for universal quantum computation. In Sec. \ref{subsubsec:fermion-sign problem} we discuss the generalization to arbitrary $n$.

\subsection{Construction of Logical Qubits} \label{subsec:construction}
We consider a 1D cloud of ``a'' atoms broken by a set of potential barriers into three segments, each of length $N$. We envision using spin-dependent potentials so that the barriers are invisible to atoms in the ``c'' state. Each segment of ``a'' atoms has a pair of Majorana edge modes. We label the three clouds as $p_1, p_2$, and $p_3$ and denote the positions of their edges by $r_1$ and $r_1'$, $r_2$ and $r_2'$, and $r_3$ and $r_3'$. We use the construction in Sec. \ref{sec:Majorana modes} to uniquely define fermionic modes $\hg_1\+, \hg_2\+$, and $\hg_3\+$ from the edge modes localized at $r_1$ and $r_1'$, $r_2$ and $r_2'$, and $r_3$ and $r_3'$.

The ground-state manifold of this system is eight-fold degenerate, $|g_{\pm\pm\pm}\rangle$. We define $|g_{---}\rangle$ as the vacuum of quasiparticles: $\hg_1|g_{---}\rangle = \hg_2|g_{---}\rangle = \hg_3|g_{---}\rangle = 0$ and define the other ground states as $|g_{\sigma\sigma'\sigma''}\rangle = \left(\hg_1\+\right)^{\frac{1+\sigma}{2}}\left(\hg_2\+\right)^{\frac{1+\sigma'}{2}}\left(\hg_3\+\right)^{\frac{1+\sigma''}{2}}|g_{---}\rangle$, where $\sigma,\sigma',\sigma''=\pm$. For example, $|g_{+++}\rangle = \hg_1\+\hg_2\+\hg_3\+|g_{---}\rangle$. We will use the notations $|g_{\sigma\sigma'\sigma''}\rangle$ and $|g_\sigma\rangle\otimes|g_{\sigma'}\rangle\otimes|g_{\sigma''}\rangle$ interchangeably, where $\otimes$ is the Cartesian product.

Of the eightfold degenerate states, four have the same parity. We construct two logical qubits from these four states and assign them the labels:
\begin{equation}\begin{array}{rlclc}
|--\rangle &\equiv& |g_{--+}\rangle &=& \hg_3\+|g_{---}\rangle \\
|-+\rangle &\equiv& |g_{-+-}\rangle &=& \hg_2\+|g_{---}\rangle \\
|+-\rangle &\equiv& |g_{+--}\rangle &=& \hg_1\+|g_{---}\rangle \\
|++\rangle &\equiv& |g_{+++}\rangle &=& \hg_1\+\hg_2\+\hg_3\+|g_{---}\rangle.
\label{eqn:construction}
\end{array}\end{equation}
We denote the logical qubits by $l_1$ and $l_2$. We call $p_1$ and $p_2$ the representational bits as their states are identical to those of the logical qubits $l_1$ and $l_2$. The ancillary bit $p_3$ serves the purpose of maintaining the total parity.

To spectroscopically measure and perform gate operations on the logical qubit, we electromagnetically excite the atoms from the ``a'' state to the ``c'' state. We find that we can measure the state of the qubits and implement the initialization, $X, Y$, and $Z$ gates on the logical qubits by addressing each physical qubit separately as outlined in Secs. \ref{sec:Gamma} and \ref{sec:gates}. To implement the Hadamard and two-qubit gates, we address two physical qubits simultaneously. In the following, we let $X_1$ denote the $X$ gate on the physical qubit $p_1$ and $X_1^{\rm{logical}}$ denote the $X$ gate on the logical qubit $l_1$. We use similar notations for the other gates. Below we describe the logical gates.

\begin{figure}[htbp]
   \unitlength=1in
   \psfragscanon
   \psfrag{omega}[][][1.5]{\begin{picture}(0,0)
    \put(-0.55,-0.1){\makebox(0,0)[l]{
    $\frac{\omega+\mu_c}{J_c}$
   }}\end{picture}}
   \psfrag{Gamma}[][][1.5]{\begin{picture}(0,0)
    \put(0.1,0.05){\makebox(0,0){
    $\frac{2J_c\hbar}{|\Omega|^2}\Gamma$
   }}\end{picture}}
  \includegraphics[width=0.9\columnwidth]{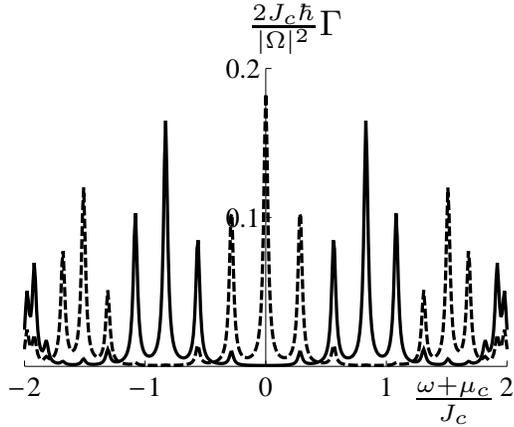}
  \caption{Absorption spectrum of physical qubit $p_1$ trapped in a linear geometry, where barriers create three physical qubits from one wire. Parameters used: $\Delta_j=J, \mu=0, N=7$. The solid curve corresponds to the spectrum of $\gplus$, and the dashed curve to the spectrum of $\gminus$.}
  \label{fig:Gamma logical qubit}
\end{figure}

\subsection{Measurement and Initialization}
As mentioned earlier, the states of the logical qubits $l_1$ and $l_2$ coincide with the states of physical qubits $p_1$ and $p_2$. The spectra of each physical qubit are given by expressions derived in Sec. \ref{sec:Gamma}. These spectra are plotted in Fig. \ref{fig:Gamma logical qubit}. Since the cloud of ``c'' atoms is three times as long as the segment corresponding to a physical qubit, the spectra contain sets of three peaks interdigitated among each other.

Projective initialization of the two logical qubits can be achieved by three single-photon transitions, one for each physical qubit. For example, to set the logical qubits in the state $|++\rangle$, we initialize each physical qubit in the state $\gplus$. To achieve this we illuminate each wire separately at the frequencies which drive the qubits into the desired state.

\subsection{Coherent Single Qubit Operations}\label{subsec:single-qubit}
We implement coherent single-qubit gates by a series of microwave pulses on individual physical qubits. One complication is that each microwave pulse not only flips a qubit, but can also add an undesired phase determined by the other qubits. This phase is a result of fermionic commutation relations. For illustration, we consider the action of the operator $\hg_2+\hg_2\+$ (which is the relevant operator when the left half of $p_2$ is illuminated, and would naively be expected to give the $X_2$ gate):
\begin{equation} \begin{array}{rlclc}
(\hg_2+\hg_2\+)|g_{-+\sigma}\rangle &=& |g_{--\sigma}\rangle &=& -Z_1X_2|g_{-+\sigma}\rangle, \\
(\hg_2+\hg_2\+)|g_{++\sigma}\rangle &=& -|g_{+-\sigma}\rangle &=& -Z_1X_2|g_{++\sigma}\rangle.
\end{array}\end{equation}
This operation yields $-Z_1X_2$ instead of the intended gate $X_2$. In the generic case, the result of a rotation $\left(z_i \hg_i + z_i^* \hg_i\+\right)$ on a state $\otimes_{j=1}^3|g_{\sigma_j}\rangle$ is
\begin{equation}\begin{split}
\left(z_i \hg_i + z_i^* \hg_i\+\right)\left(\otimes_{j=1}^3|g_{\sigma_j}\rangle\right) = \left(\otimes_{j<i}-Z_j|g_{\sigma_j}\rangle\right)\otimes\\
    \left(\left(z_i \hg_i + z_i^* \hg_i\+\right)|g_{\sigma_i}\rangle\right)\otimes\left(\otimes_{j>i}|g_{\sigma_j}\rangle\right)
\end{split}\end{equation}
The sequence of $-Z$ gates above can be thought of as a Jordan-Wigner transformation. As explicitly shown below, a simple sequence of pulses can remove these unwanted phases for the $X, Y$ and $Z$ gates.

Incorporating pulses to remove unwanted phases, we find
\begin{equation}\begin{array}{rll}
 X^{\rm{logical}}_1 &=& X_1X_3 \\ &=& (\hg_1-\hg_1\+)(\hg_2+\hg_2\+)(\hg_2-\hg_2\+)(\hg_3+\hg_3\+), \\
 Y^{\rm{logical}}_1 &=& Y_1X_3 \\ &=& (\hg_1+\hg_1\+)(\hg_2+\hg_2\+)(\hg_2-\hg_2\+)(\hg_3+\hg_3\+), \\
 X^{\rm{logical}}_2 &=& X_2X_3 \\ &=& (\hg_2-\hg_2\+)(\hg_3+\hg_3\+), \\
 Y^{\rm{logical}}_2 &=& Y_2X_3 \\ &=& (\hg_2+\hg_2\+)(\hg_3+\hg_3\+).
\end{array} \label{eqn:logicalXYZ} \end{equation}
As in Sec. \ref{sec:gates}, each operator in parentheses corresponds to illuminating one part of one segment. As before, $Z^{\rm{logical}}_i = -i X^{\rm{logical}}_iY^{\rm{logical}}_i$.

The logical Hadamard gate, $H^{\rm{logical}}_i$, requires two-qubit operations (one representational bit and one ancillary bit) and is discussed in Sec. \ref{subsubsec:H and CY} along with the $C^Y$ gate.

\subsection{Two-Qubit Operations} \label{subsec:2-qubit}
We achieve two-qubit operations on physical qubits $p_i$ and $p_j$ by simultaneously illuminating $p_i$ and $p_j$ with a microwave coupling strength $\Omega<J_c/N$. To achieve different rotations in the ground-state manifold, we focus microwave pulses of different frequencies, durations and spatial distributions. For concreteness, we consider the rotation given by $\hg_1+\hg_1\++\hg_2+\hg_2\+$. To achieve this, we illuminate the left half of $p_1$ and $p_2$ simultaneously. Labeling the ``c'' eigenstate being excited by $k$, the Hamiltonian governing the excitation to that mode has the form
\begin{equation}\begin{split}
 \hat{H}_{\rm{2qubit}}^{\rm{res}} = \sum_{0\leq j<N/2} f_1(r_1+j)\psi_k(r_1+j)\hc_k\+\frac{\hat\gamma_1+\hat\gamma_1\+}{2}
   \\+ f_2(r_2+j) \psi_k(r_2+j)\hc_k\+\frac{\hat\gamma_2+\hat\gamma_2\+}{2} + f_1(r_1'-j) \psi_k(r_1'-j)\\ \times\hc_k\+\frac{\hat\gamma_1-\hat\gamma_1\+}{2}
   + f_2(r_2'-j)\psi_k(r_2'-j)\hc_k\+\frac{\hat\gamma_2-\hat\gamma_2\+}{2} + h.c
 \label{eqn:H2qubit}
\end{split}\end{equation}
Since $f_1(r_1+j)$ and $f_2(r_2+j)$ exponentially decay away from $j=0$, we neglect the coefficients of $\hg_1-\hat{\gamma_1}\+$ and $\hg_2-\hat{\gamma_2}\+$. We find that the microwave frequency should be chosen such that $\sum_{0\leq j<N/2}f_1(r_1+j)\psi_k(r_1+j) = \sum_{0\leq j<N/2}f_2(r_2+j)\psi_k(r_2+j)$ whose value we denote by $\alpha_k$. In that case, Eq. (\ref{eqn:H2qubit}) reduces to
\begin{equation}
\hat{H}_{\rm{2qubit}}^{\rm{res}} = \Omega \frac{\alpha_k\hc_k\+-\alpha_k^*\hc_k}{2}\left(\hat\gamma_1+\hat\gamma_1\+ + \hat\gamma_2+\hat\gamma_2\+\right).
\end{equation}
A $\pi$ pulse lasting $T=\frac{\pi\hbar}{|\alpha_k|\Omega}$ performs the intended operation $\hat\gamma_1+\hat\gamma_1\+ + \hat\gamma_2+\hat\gamma_2\+$. Other rotations in the ground-state manifold can be produced by choosing the frequency, duration, and spatial distribution of the microwave pulse appropriately.

Below we provide explicit algorithms to perform the $H^{\rm{logical}}_i$ and $C^Y$ gates. We first introduce two operations that will be useful building blocks and construct the $H^{\rm{logical}}_i$ and $C^Y$ gates out of these building blocks. In Sec. \ref{subsubsec:fermion-sign problem}, we consider the action of generic rotations on the qubits and the problems encountered in generic rotations.

\subsubsection{Gates Involving Two-Qubit Operations} \label{subsubsec:H and CY}
The first operation we consider, denoted by $H_{ij}$, consists of two steps which perform the rotation $(\hg_i\+ + \hg_i + \hg_j\+ + \hg_j)(\hg_i - \hg_i\+)$. The second operation we introduce also consists of two steps: $S_{ij}=(\hg_i + \hg_i\+ + \hg_j + \hg_j\+)(\hg_i - \hg_i\+ + \hg_j - \hg_j\+)$.
We construct $H^{\rm{logical}}_i$ and $C^Y$ gates out of $H_{ij}, S_{ij}$ and single-qubit gates.

In terms of these building blocks,
\begin{equation}\begin{array}{rll}
H_2^{\rm{logical}} &=& H_{23}, \\
C_{12}^Y &=& Z_1Z_2S_{23}, \\
C_{21}^Y &=& Z_1S_{12}S_{23}S_{12},\ \rm{and}\\
H_1^{\rm{logical}} &=& C_{21}^YS_{13}.
\end{array}\end{equation}

\subsubsection{Entanglement in Two-Qubit Operations} \label{subsubsec:fermion-sign problem}
In this section we explore generic two-qubit operations that can be performed by microwave pulses on pairs of qubits. The $H_{ij}$ and $S_{ij}$ operations introduced earlier are special cases.

The term in the Hamiltonian [Eq. (\ref{eqn:HMW})] which resonantly couples the Majorana mode to a ``c'' eigenstate labeled by $k$ has the form
\begin{equation}
 \hat{H}^{\rm{res}} = \Omega(\hc_k\+\hat{h}_{ij} + \hat{h}_{ij}\+\hc_k),
\end{equation}
where $\hat{h}_{ij}$ consists of edge mode operators. The most generic $h_{ij}$ is Hermitian; i.e, it has the form:
\begin{equation}
 \hat{h}_{ij} = z_i\hg_i + z_i^*\hg_i\+ + z_j\hg_j + z_j^*\hg_j\+. \label{eqn:general 2-qubit rotation}
\end{equation}
This can be derived as follows. In order for the ``c'' atoms to disentangle from the rotation performed on the qubits, $\hat{h}_{ij}$ has to satisfy $\hat{h}_{ij}\+ = e^{i\phi}\hat{h}_{ij}$ where $\phi$ is an arbitrary phase. By absorbing this phase into the ``c'' operators, we arrive at the form in Eq. (\ref{eqn:general 2-qubit rotation}).

We illustrate that $\hat{h}_{ij}$ leads to different physical consequences when acting upon neighboring qubits or distantly spaced qubits. If the segments $i$ and $j=i+1$ are neighbors, the result of the generic pulse in Eq. (\ref{eqn:general 2-qubit rotation}) only involves the physical qubits on those sites, and
\begin{equation}\begin{split}
 \hat{h}_{i,i+1}|g_{\sigma_i\sigma_{i+1}}\rangle = \left(z_i\hg_i + z_i^*\hg_i\+|g_{\sigma_i}\rangle\right)\otimes|g_{\sigma_{i+1}}\rangle\\ + (-1)^{\frac{1+\sigma_i}{2}} |g_{\sigma_{i}}\rangle\otimes\left(z_{i+1}\hg_{i+1} + z_{i+1}^*\hg_{i+1}\+|g_{\sigma_{i+1}}\rangle\right),
\end{split}\end{equation}
where $\sigma_i,\sigma_{i+1}=\pm$. If instead the segments are further spaced, the action of the pulse involves all intervening qubits. For example, if they are separated by one segment (which is the largest separation allowed for three physical qubits),
\begin{equation}\begin{split}
  \hat{h}_{i,i+2}|g_{\sigma_i\sigma_{i+1}\sigma_{i+2}}\rangle = \left(z_i\hg_i + z_i^*\hg_i\+|g_{\sigma_i}\rangle\right)\otimes|g_{\sigma_{i+1},\sigma_{i+2}}\rangle \\ +(-1)^{\frac{2+\sigma_i+\sigma_{i+1}}{2}} |g_{\sigma_{i},\sigma_{i+1}}\rangle\otimes\left(z_{i+2}\hg_{i+2} + z_{i+2}^*\hg_{i+2}\+|g_{\sigma_{i+2}}\rangle\right).
\end{split}\end{equation}
The generalization to longer chains is straightforward.

The extra phases produced when $h_{ij}$ acts on nonadjacent qubits are the reason why the implementation of $H_1^{\rm{logical}}$ and $C_Y^{21}$ in Sec. \ref{subsubsec:H and CY} were so much more complicated than  $H_2^{\rm{logical}}$ and $C_Y^{12}$. Due to these phases, extending our proposal to more than two logical qubits is non-trivial. However we see no impediment to constructing generic gates for chains of $N$ qubits for arbitrary $N$.

\section{Summary} \label{sec:Summary}
In summary, we have presented an experimentally feasible method to perform quantum computing with Majorana fermions in cold gases using microwaves. We considered two geometries which give rise to Majorana fermion excitations: a 2D array of coupled 1D wires, and a single 1D wire embedded in a superfluid cloud. We proposed various methods to generate nearest-neighbor interactions between atoms in the coupled wires, which is crucial to creating Majorana fermions. We modeled these systems with a mean-field theory and studied their single-particle excitation spectra. We observed that the systems supported Majorana modes for a certain range of parameters, and that in this ``topologically non-trivial phase'', the ground state is doubly degenerate. These two degenerate states can be used as a qubit. We calculated the absorption spectra, and showed that the lineshape gives evidence of Majorana fermions and can be used to measure the state of the qubit. We further showed that absorption of a photon flips between the degenerate states. We proposed that this feature could be used to perform quantum gates on the qubit. We found the geometries for which this protocol works. We presented algorithms to perform certain quantum gates on individual physical qubits. We constructed logical qubits out of physical qubits and gave generic arguments to perform rotations of the qubits. In addition to these arguments, we gave explicit pulse sequences for constructing a universal set of quantum gates for two logical qubits encoded in three physical qubits, hence allowing our system to be used for universal quantum computation.

\section*{ACKNOWLEDGEMENTS}
We would like to thank Matthew Reichl for useful discussions. This work is supported by the National Science Foundation Grant no. PHY-1068165 and a grant from the Army Research Office with funding from the DARPA OLE program.

\appendix
\section{Superfluid gap from nearest-neighbor interactions} \label{sec:Appendix}
Nearest-neighbor interactions in the tight-binding model [Eq. (\ref{eqn:full Hsys})] for the ``a'' atoms in an optical lattice lead to a superfluid gap $\Delta$, given implicitly by the gap equation,
\begin{equation}
 \frac{-1}{V} = \frac{1}{N}\sum_k \frac{\sin^2ka}{\sqrt{(2J\cos ka-\mu)^2+(2\Delta\sin ka)^2}},
 \label{eqn:gap equation}
\end{equation}
where $N$ is the number of lattice sites, $a$ is the lattice constant, $V$ is the nearest-neighbor interaction strength, $J$ is the hopping amplitude, and allowed momenta $k=\frac{2n\pi}{Na},\ (n=0,1,..N-1)$ are summed over. $J$ is typically controlled by the depth of the optical lattice, and $V$ depends on the parameters dictating the mechanism creating nearest-neighbor interactions, and the $s$-wave scattering length if applicable. Below we calculate the superfluid gap created in three different mechanisms producing nearest-neighbor interactions. 

\subsection{Dipolar molecules} \label{subsec:dipoles}
Dipolar molecules such as KRb or LiCs have nearest-neighbor dipole-dipole interactions. The interaction strength of two molecules at adjacent lattice sites, with dipole moments $\vec{d}_1$ and $\vec{d}_2$ is
\begin{equation}
 V = \frac{\vec{d}_1\cdot\vec{d}_2-3\vec{d}_1\cdot\hat{r}\vec{d}_2\cdot\hat{r}}{4\pi\epsilon_0a^3},
\end{equation}
where $a$ is the lattice constant and $\hat r$ is the unit vector joining the two lattice sites. For typical dipole moments of the order of 10 D and typical lattice spacings of a few $\mu$m, this interaction strength is on the order of a few kHz. For hopping amplitudes of a few kHz, Eq. (\ref{eqn:gap equation}) results in a superfluid gap of the order of kHz.

\subsection{Artificial spin-orbit coupling} \label{subsec:SOC}
Nearest-neighbor interaction arises in spin-orbit coupled gases as an effective interaction between atoms in dressed states (the helicity states). The Hamiltonian describing the bare atoms is
\begin{equation}\begin{split}
 \hH =
\sum_k \left(\begin{array}{cc} \ha_{\uparrow k+k_l}\+ & \ha_{\downarrow k-k_l}\+ \end{array}\right)
\left(\begin{array}{cc}\epsilon_{k+k_L} & \Omega \\ \Omega & \epsilon_{k-k_L} \end{array}\right)
\left(\begin{array}{c} \ha_{\uparrow k+k_l} \\ \ha_{\downarrow k-k_l} \end{array}\right) \\
+ V\sum_i \ha_{\uparrow i}\+ \ha_{\downarrow i}\+ \ha_{\downarrow i} \ha_{\uparrow i}
\end{split}\end{equation}
where the Raman coupling strength $\Omega$ will dictate the energy gap between the two helicity states, $\epsilon_k = -2J\cos ka$ is the kinetic energy due to tunneling in an optical lattice. For small momenta, the spin-orbit coupling strength is $-2Ja\sin k_La$, where $k_L$ is the recoil momentum due to a Raman photon. The effective nearest-neighbor interaction  between two atoms in the lower helicity state is
\begin{equation}
 V_{eff} \simeq V \left(\frac{2J}{\Omega}\sin k_La\right)^2
\end{equation}
for $\Omega >> 2J$. $J_{eff}\simeq J\cos k_La$ is the effective hopping amplitude between atoms in dressed states. For typical scattering lengths of the order of 100 Bohr radii at Feshbach resonance and typical hopping amplitudes of a few kHz, the superfluid gap from Eq. (\ref{eqn:gap equation}) is a few kHz.

\subsection{Spin-dependent lattices}\label{subsec:Spin-dependent lattices}
\begin{figure}[htbp]
   \includegraphics[width=0.5\columnwidth]{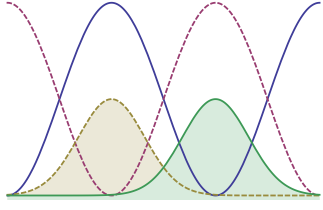}
   \caption{(Color online) Schematic of a deep spin-dependent lattice and a transverse magnetic field.} \label{fig:spin-dependent lattice}
\end{figure}
Another way to create strong nearest-neighbor interactions between atoms is to increase the overlap between Wannier functions at two adjacent sites on an optical lattice. For example, consider the spin-dependent lattice along the $x$ direction illustrated in Fig. \ref{fig:spin-dependent lattice}, with its spin-quantization axis along the $z$ direction. The sites for $\uparrow$-spins are shifted by half a lattice constant relative to the sites of the $\downarrow$-spins. Equivalently, we can consider a superlattice with $\uparrow$-spins on the even sites and $\downarrow$-spins on the odd sites. In the presence of a transverse magnetic field $B\hat y$, this system can be modeled by the Hamiltonian
\begin{widetext}
\begin{eqnarray}
 \hat{H} &=& \int dx\, \frac{-\hbar^2}{2m}\sum_\sigma\left(\hat\psi\+_\sigma(x)\frac{\partial^2}{\partial x^2}\hat\psi_\sigma(x)\right)
 + U_0\cos^2\frac{\pi x}{a}\hat\psi\+_\uparrow(x)\hat\psi_\uparrow(x) + U_0\sin^2\frac{\pi x}{a}\hat\psi\+_\downarrow(x)\hat\psi_\downarrow(x) 
 \\&&\qquad\nonumber
 -\mu_0B(\hat\psi\+_\uparrow(x)\hat\psi_\downarrow(x)+\hat\psi\+_\downarrow(x)\hat\psi_\uparrow(x)) + g\hat\psi\+_\uparrow(x)\hat\psi\+_\downarrow(x)\hat\psi_\downarrow(x)\hat\psi_\uparrow(x),
 \label{eqn:H spin-dependent lattice}
\end{eqnarray}
\end{widetext}
where $m$ is the mass of the fermion, $U_0$ is the lattice depth, $\mu_0$ is the Bohr magneton, and $g$ is the on-site interaction strength. Here $\hat\psi_\uparrow(x) = \sum_i \hat{a}_{2i\uparrow}\phi(x-x_{2i})$ and $\hat\psi_\downarrow(x) = \sum_i \hat{a}_{2i+1\downarrow}\phi(x-x_{2i+1})$ are field operators for $\uparrow$-spins and $\downarrow$-spins, where $\hat{a}_{i\sigma}$ annihilates a fermion with spin $\sigma$ at site $i$, $\phi(x)$ is the Wannier function, $x_i$ are the positions of the lattice sites, and all sites $i$ have been summed over. The first term in Eq. (\ref{eqn:H spin-dependent lattice}), which gives rise to hopping between sites with the same spin on the lattice, can be quenched by increasing the lattice depth. The magnetic field enables the fermions to hop between adjacent sites on the superlattice while flipping their spin. The last term in Eq. (\ref{eqn:H spin-dependent lattice}) gives rise to interactions between adjacent fermions with opposite spin. This is clearly illustrated by rewriting the Hamiltonian in terms of new operators $\hat{b}_{2i}=\hat{a}_{2i\uparrow}, \hat{b}_{2i+1}=\hat{a}_{2i+1\downarrow}$ as
\begin{equation}
 \hat{H} = \sum_i -J_{eff}(\hat{b}\+_i\hat{b}_{i+1}+h.c) + V_{eff} \hat{b}\+_i\hat{b}\+_{i+1}\hat{b}_{i+1}\hat{b}_i.
\end{equation}
This is equivalent to a model for a gas of spinless fermions with nearest-neighbor interactions on an optical lattice. Direct computation yields
\begin{equation}
 J_{eff} = \mu_0B \int dx \phi^*(x)\phi(x-a/2)
\end{equation}
and
\begin{equation}
 V_{eff} = g \int dx |\phi(x)|^2|\phi(x-a/2)|^2.
\end{equation}
Magnetic fields of a few Gauss lead to hopping amplitudes in the MHz range. If we tune the interaction strength to a few MHz via a Feshbach resonance, the superfluid gap from Eq. (\ref{eqn:gap equation}) would be on the order of a few MHz.

\section{Edge modes} \label{sec:Appendix2}
The creation operator for the zero-energy quasiparticle has the form
\begin{equation}
 \hmaj\+ = \sum_j f_0(j)\left(\frac{\ha_j\++\ha_j}{2}+i\frac{\ha_{N+1-j}\+-\ha_{N+1-j}}{2i}\right).
\end{equation}
Demanding that the quasiparticle created by this operator has zero energy leads to difference equations governing the coherence factors $f_0(j)$,
\begin{equation}\begin{split}
 (J-\Delta_{j-1})f_0(j-1) + (J+\Delta_{j+1})f_0(j+1) - \mu f_0(j) = 0,\\ 1<j<N,
\end{split}\end{equation}
For the non-self-consistent case of uniform $\Delta_j=\Delta$, these equations are solved by assuming a solution of the form $f_0(j) = \alpha x_+^j + \beta x_-^j$. $\alpha$ and $\beta$ are determined by the boundary conditions
\begin{equation}\begin{array}{rllcl}
 f_0(2) &=& \frac{\mu}{J+\Delta}f_0(1),\ &\rm{if}& \Delta>0,\\
 f_0(N-1) &=& \frac{\mu}{J-\Delta}f_0(N),\ &\rm{if}& \Delta<0,
\end{array}\end{equation}
and the normalization condition
\begin{equation}
 \sum_j \left|\frac{f_0(j)+f_0(N+1-j)}{2}\right|^2 + \left|\frac{f_0(j)-f_0(N+1-j)}{2}\right|^2 = 1.
\end{equation}
The difference equations yield
\begin{equation}
 x_\pm = \frac{\mu}{2(J+\Delta)} \pm \sqrt{\left(\frac{\mu}{2(J+\Delta)}\right)^2+\frac{\Delta-J}{\Delta+J}},
\end{equation}
and the boundary conditions yield $\alpha=-\beta$ if $\Delta>0$, and $\alpha=-\left(\frac{x_-}{x_+}\right)^{N+1}\beta$ if $\Delta<0$.

\bibliography{references}

\end{document}